\documentclass
[twocolumn,showpacs,amssymb,prl,aps,superscriptaddress,10pt]{revtex4-1}

\usepackage{epsfig}
\usepackage{verbatim}
\usepackage{dcolumn}
\usepackage{bm}

\def\apj{ApJ}                 
\def\apjl{ApJL}                
\def\aap{A\&A}                
\def\mnras{MNRAS}             
\def\pasj{PASJ}               
\def\physrep{Phys.~Rep.}   
\def\jcap{J. Cosmology Astropart. Phys.}
\def\araa{ARA\&A}             

\newcommand{\kpc}{\mbox{ kpc}}
\newcommand{\Mpc}{\mbox{ Mpc}}
\newcommand{\se}{\mbox{ s}}
\newcommand{\km}{\mbox{ km}}

\newcommand{\muG}{\mbox{ $\mu$G}}
\newcommand{\grad}{\bm{\nabla}}

\newcommand{\ie}{\emph{i.e.} }
\newcommand{\eg}{\emph{e.g.,} }

\newcommand{\ave}[1]{{\overline{#1}}}

\newcommand{\mach}{{\mathcal{M}}}
\newcommand{\till}{{\mbox{--}}}

\newcommand{\rCF}{{r_{\mbox{\scriptsize{cf}}}}}
\newcommand{\Ptot}{{P_{\mbox{\scriptsize{tot}}}}}

\begin{document}

\title{Strong Magnetization Measured in the Cool Cores of Galaxy Clusters}

\author{Ido Reiss}

\email{reissi@post.bgu.ac.il}
\affiliation{Physics Department, Ben-Gurion University of the Negev, POB 653, Be'er-Sheva 84105, Israel}
\affiliation{Physics Department, Nuclear Research Center Negev, POB 9001, Be'er-Sheva 84190, Israel}

\author{Uri Keshet}
\affiliation{Physics Department, Ben-Gurion University of the Negev, POB 653, Be'er-Sheva 84105, Israel}

\date{\today}

\begin{abstract}
Tangential discontinuities, seen as X-ray edges known as cold fronts (CFs), are ubiquitous in cool-core galaxy clusters. We analyze all 17 deprojected CF thermal profiles found in the literature, including three new CFs we tentatively identify (in clusters A2204 and 2A0335). We discover small but significant thermal pressure drops below all nonmerger CFs, and argue that they arise from strong magnetic fields below and parallel to the discontinuity, carrying $10\%-20\%$ of the pressure. Such magnetization can stabilize the CFs, and explain the CF---radio minihalo connection.
\end{abstract}


\maketitle

In the past decade, high resolution X-ray images revealed an abundance of density and temperature discontinuities known as cold fronts (CFs). They are broadly classified as cool-core (CC) CFs vs. merger CFs (for a review, see \cite{markevitch07}).
Here we focus exclusively on CC CFs.

Such CFs are observed in most of the otherwise relaxed, CC clusters \cite{Markevitch03}, inside the core and sometimes beyond it. They are usually nearly concentric or spiral, and multiple CFs are often observed in the same cluster. The plasma beneath the CF is typically denser, colder, lower in entropy, and higher in metallicity, than the plasma above it. The temperature $T$ contrast across such a CF is \citep{owers09} $T_{o}/T_{i} \sim 2$, where inside/outside subscripts $i/o$ refer to regions closer to/farther from the cluster center, or equivalently below/above the CF.

Such CFs are thought to be a quasi-spiral tangential discontinuity surface seen in projection \cite{ascasibar06, Keshet12}.
They may reflect large-scale ``sloshing'' oscillations of the intracluster medium (ICM), driven by mergers \cite{Markevitch01,
Tittley05,ascasibar06}, or feedback from the central active galactic nucleus (AGN)
\cite{Churazov03, fujita04}. They were also proposed to be the signature of long-lived spiral bulk flows underlying cool cores
\cite{Keshet12}.

Deprojected thermal profiles across core CFs reveal the presence of fast, nearly sonic flows beneath CFs, and tangential shear layers extending below them \cite{Keshet10}.
Such shear can produce the strong magnetization needed to stabilize the CF against Kelvin-Helmholtz instabilities (KHI), $\eta_B\equiv P_B/\Ptot = B^2/(8\pi \Ptot)= 10\till20\%$, if the shear layer is sufficiently long and narrow \cite{Keshet10}, as expected for example in a long spiral CF.
Here, $B$ and $P_B$ are the magnetic amplitude and pressure, and $\Ptot$ is the total pressure.
Such shear-induced magnetization levels were indeed reproduced in sloshing simulations \cite{zuhone11}.

There is circumstantial evidence for strong magnetic fields along and below CFs. The observed thinness of the discontinuity, much narrower than the Coulomb mean free path, requires a magnetic suppression of transport across the CF \cite{vikhlinin01,ettori00}. The CF--radio minihalo connection \cite{mazzotta08,giacintucci11,zuhone12}, in particular when combined with the radio--X-ray correlation in mini-halos, suggests strong magnetization below CFs, exceeding the cosmic microwave background equivalent $B\sim3\muG$ \cite{keshetloeb10}.
Previous measurements of core magnetic fields \cite{carilli02,clarke04,govoni10,Bonafede11} produced a wide range of results, but are not specific to CF regions.

In this \emph{letter} we report the discovery of a discontinuity in the thermal pressure $P_{th}$ across core CFs. We interpret this as a discontinuous jump in \emph{nonthermal} pressure $P_{nt}$ below the CF, and argue that it must be predominantly magnetic. We assume a Hubble constant of $H_0=70\km\se^{-1}\Mpc^{-1}$, and a $77\%$ hydrogen mass fraction. Error bars represent $1\sigma$ confidence levels.

\emph{Core CFs do show a thermal pressure discontinuity.---} Unlike merger CFs, core CFs were so far thought to involve no change in $P_{th}$, \ie appear thermally isobaric (see \cite{markevitch07} and references therein). To critically examine this, we analyze the thermal profiles near core CFs, and derive the pressure on both sides of the discontinuity.

\begin{figure}[ht!]
\centerline{\epsfxsize=8.5cm \epsfbox{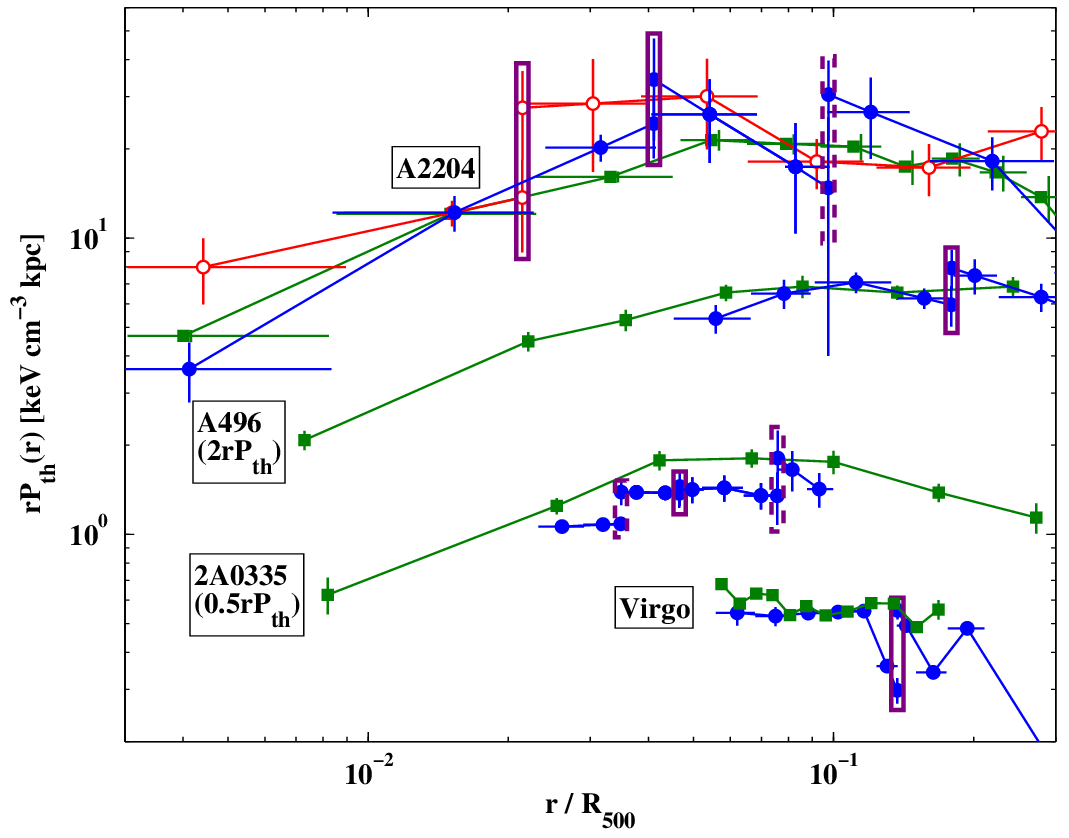}}
\caption{\label{fig:profiles}
Deprojected $r P_{th}(r)$ profiles (circles, empty or filled for different sectors, including $\rCF$ extrapolation) across CFs (rectangles; dashed for putative CFs), and azimuthal CC averages (squares). Lines are guides to the eye.
For visibility, we show $2rP_{th}$ in A496, and $0.5rP_{th}$ in 2A0335.
In all but A2204, the mean $P_{th}$ was divided by a fudge factor $\sim1.7$ in order to match the CF sectors away from $\rCF$; this has no effect on our results.
Mean profile references: A496 and 2A0335 \cite{tamura03}, Virgo \cite{roediger11}, and A2204 \cite{sanders09a2204}. CF profiles: A2204 \cite{Sanders05}, A496 \cite{Tanaka06}, 2A0335 \cite{Sanders09} and Virgo \cite{urban11}.
}
\end{figure}

We extract all thermal profiles across core CFs in the literature, and select only profiles that have been deprojected along the line of sight. The underlying assumptions of a single-phase ideal gas, spherically symmetrically distributed near the CF, are discussed below. We find 14 such CFs in 10 CCs, where $P_{th}$ can be estimated on both sides of the CF. We compute $P_{th}$ based on the deprojected profiles of $T$ and particle number density $n$. Some pressure profiles are shown in Figure \ref{fig:profiles}, using $rP_{th}(r)$ to highlight deviations from the $P\sim r^{-1}$ profile.

\begin{figure}[t!]
\centerline{\epsfxsize=8.5cm \epsfbox{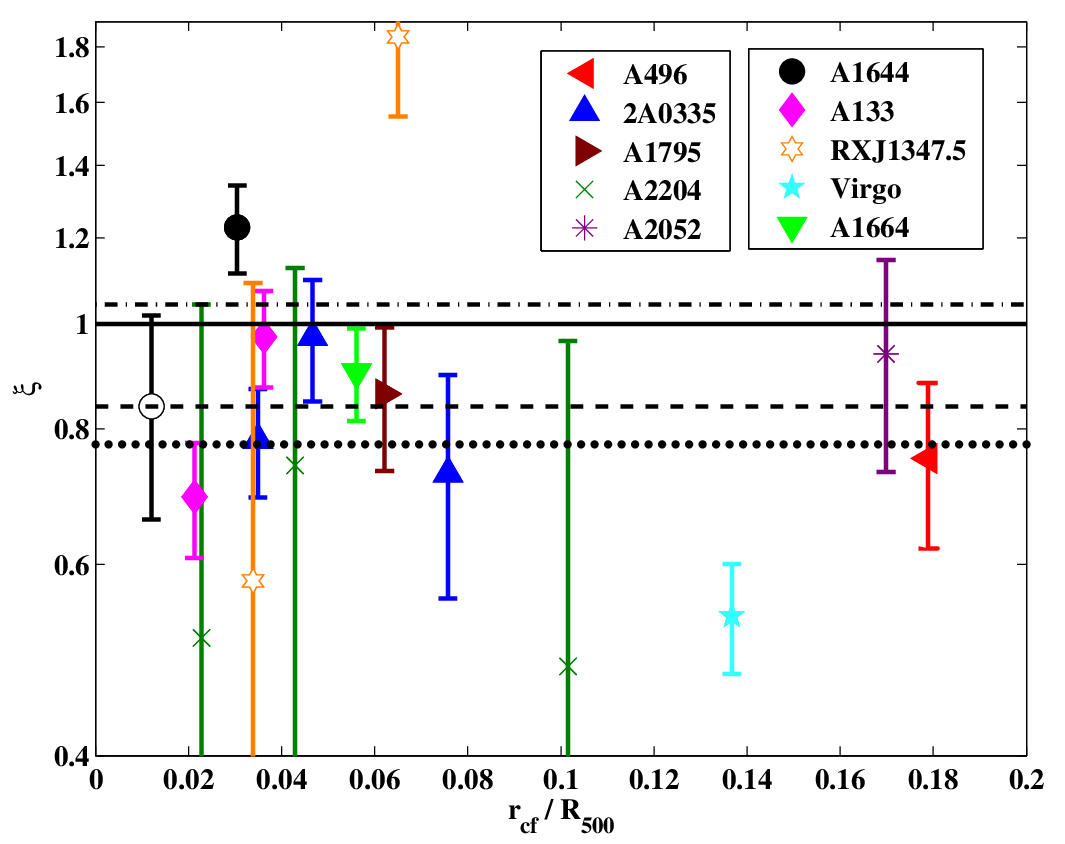}}
\caption{\label{fig:ratios}
Deprojected thermal pressure jump $\xi=P_i/P_o$ vs. normalized radius $r/R_{500}$, for all our CFs. The solid $\xi=1$ line corresponds to a thermally isobaric CF. Horizontal lines show the averages of all CFs (dotted line), of our final CF sample (dashed, Eq.~(\ref{eq:mean xi})), and of the non-CF sample (dot-dashed); see text. An isothermal profile is assumed below CFs marked by empty symbols. Data references: A1664 \cite{kirkpatrick09}, A1795 \cite{Markevitch01}, A2052 \cite{dePlaa10}, A1644S \cite{Johnson10}, A133 \cite{randall10}, RXJ1347.5 \cite{johnson12}, $R_{500}$ \cite{reiprich02,urban11,babyk12,pimbblet06}.}
\end{figure}

Figure \ref{fig:ratios} shows the ratio $\xi\equiv P_i/P_o$ between $P_{th}$ just inside $(P_i)$ and just outside $(P_o)$ each CF, as a function of the CF radius $\rCF$ normalized by $R_{500}$, the radius enclosing a mean density $500$ times the critical density of the Universe. Data references are provided in the caption.

We compute $P_i$ and $P_o$ by approximating $n$ and $T$ near $\rCF$ as broken power-laws, estimated on each side of the CF from the two nearest bins. In the three cases with only one temperature bin below the CF (RXJ1347.5 and the inner CF in A1644S; empty symbols in Figure \ref{fig:ratios}), the inner gas is assumed isothermal. We tentatively identify three new CFs (dashed rectangles in Figure \ref{fig:profiles}), seen coincidentally in emission, temperature, and sometimes also metallicity: at $\rCF$ $\sim 28$ and $\sim 62 \kpc$ Southwest of 2A0335 \cite{Sanders09}, and $\rCF\sim 127\kpc$ West of A2204 \cite{Sanders05}.

Confidence levels are estimated by error propagating the temperature and density uncertainties, assumed for simplicity to be statistical, normally distributed, and uncorrelated. Our error estimates are conservative; accounting for correlations between bins should reduce the uncertainties in the inferred pressure jumps.

All but two CFs (the exceptions are discussed below) show $\xi<1$, suggesting a $P_{th}$ deficit below the CF. The statistical significance per CF is modest (from $0.3\sigma$ to $3.6\sigma$, except Virgo with $7.3\sigma$), but the accumulated evidence is significant. The $P_{th}$ deficit is already resolved, with no extrapolation to the CF, in five cases (two CFs in 2A0335, two in A2204 and in Virgo). In most CFs, it is clearly seen in the $rP_{th}$ profile (\eg Figure \ref{fig:profiles}).

\begin{figure}
\centerline{\epsfxsize=8.5cm \epsfbox{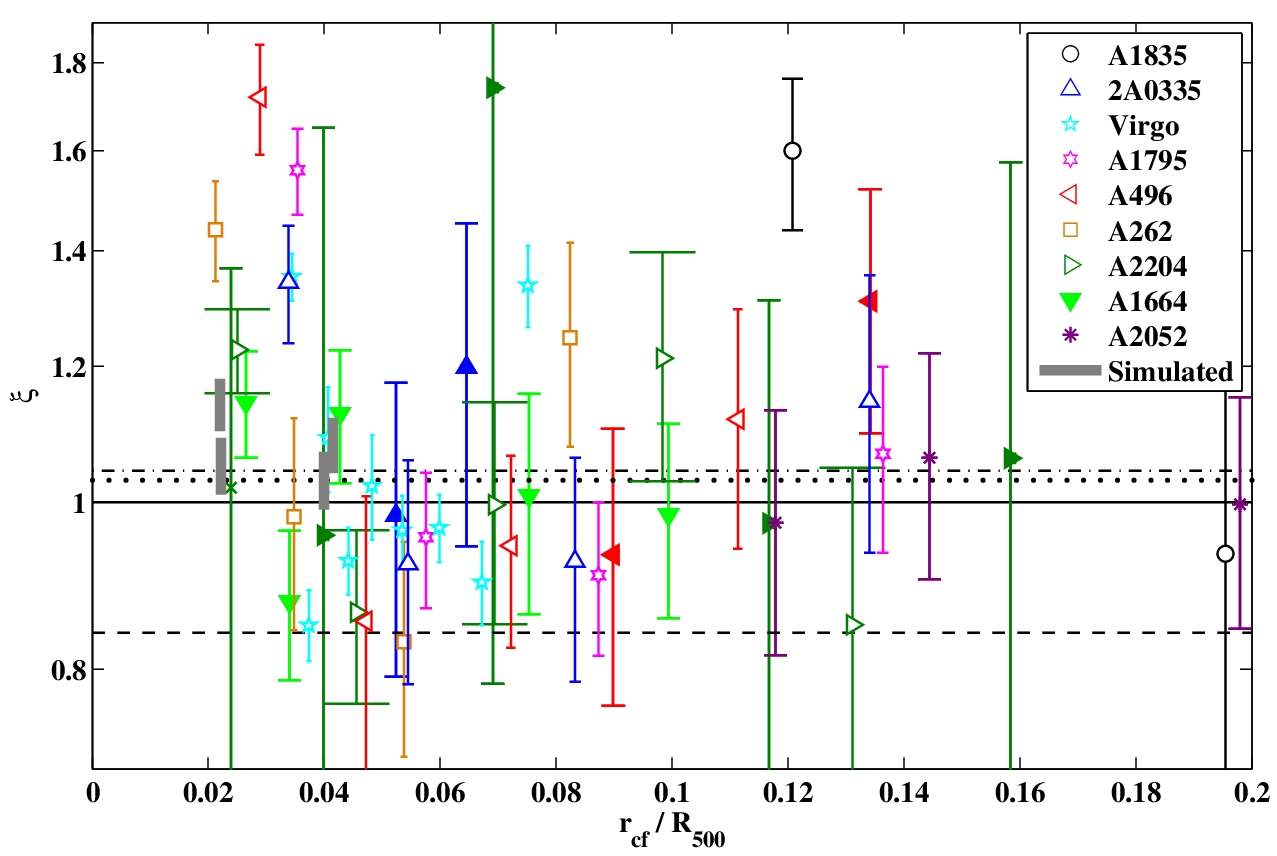}}
\caption{\label{fig:control profiles}
Control samples: fictitious CFs (filled symbols), azimuthal averages (empty symbols), and simulated isobaric CFs (no symbols; uncertainties due to CF localization). Horizontal lines show the mean $\xi$ for real (solid) vs. fictitious CFs with (dotted) and without (dot-dashed) Virgo.
}
\end{figure}

The CFs mean is $\ave{\xi}=0.77\pm0.03$. The aforementioned $\xi>1$ cases are the outer CFs in A1644S and RXJ1347.5, both undergoing a merger. These CFs appear affected by the merger; excluding them slightly changes the mean to  $\ave{\xi}=0.73\pm0.03$. The Virgo sector is very narrow and its data suggest substructure (see Figures \ref{fig:control profiles}, \ref{fig:profiles}), so are excluded as well. The mean of the remaining 14 CFs is
\begin{equation}
\label{eq:mean xi}
\ave{\xi}=0.81\pm0.03 \, .
\end{equation}
Excluding also the 3 tentatively identified CFs would give $\ave{\xi}=0.82\pm0.04$. We conclude that $\ave{\xi}\simeq 0.8$, with a large dispersion among CFs, significantly deviates below unity.

\emph{Robustness.---}
These results rely on spherical deprojection, fitting the data by a three-dimensional model of shells parallel to the CF.
This well-established method is widely applied to X-ray clusters \cite{Markevitch00,russell08,allen01}.
Due to the steep density gradients, the $\propto n^2$ emission is dominated by the edge-on shell, rendering deprojection a small correction; deprojection errors are an even smaller, second-order effect \cite{Markevitch00}.
Such measurements are sufficiently robust to detect a $20\%$ pressure jump even in a single galaxy group CF \cite{lal13, gastaldello13}; the estimated errors are small \cite[\eg][]{gastaldello13, GrahamEtAl08, blanton11}.

Systematic effects may bias $\xi$, mainly \emph{raising} it due to pressure curvature. Other effects, such as CF asphericity or $\rCF$ errors, may offset $\xi$ in a given CF. However, in our heterogeneous sample, with various projections, the main effect is an enhanced dispersion, whereas a resulting offset in $\ave{\xi}$ remains small and positive.
Aspherical CFs cannot mimic our effect, as the small $n$ contrast (median $q=1.2$) renders this a $5\%$ effect; moreover, the viewing-angle-averaged bias to $\xi$ is positive (few percent).
Most (8/14) $\rCF$ value are derived as a model parameter; no systematic difference is found in CFs located by eye.

As a general test of systematics, including deprojection and fit errors, we compute $\xi$ with the same data and in the same method used to produce Figure \ref{fig:ratios}, but for radii showing no CF. This is possible in 4 sectors, in which we place fictitious CFs between consecutive radial bins away from the true CFs. The control $\xi_c$ values thus obtained, shown in Figure \ref{fig:control profiles}, all lie above the real CF mean (Eq.~\ref{eq:mean xi}); their average is $\ave{\xi}_c=1.03\pm0.06$ (if the oscillatory profile of Virgo is included as a fifth sector, $\ave{\xi}_c=1.05\pm0.03$).
A similar analysis of deprojected, azimuthally-averaged radial profiles \cite{tamura03,sanders09a2204,roediger11} yields $\ave{\xi}_c=1.04\pm0.01$; see Figure \ref{fig:control profiles}. Such $\xi>1$ values arise from the bias discussed above.

To further test for false positives and projection effects, we repeat the analysis for the isobaric CFs in magnetic-free sloshing simulations \cite{roediger11,ascasibar06}. We bin mock CF sectors in different projections, deproject the SB, and use the projected $T$, found to be nearly isothermal on each side of these CFs. All such CFs show $\xi>1$; see Figure \ref{fig:control profiles}. No false positive is found, and projection effects are small.

A thin, unresolved shear layer near the CF cannot mimic the signal because (i) such jumps are resolved (with no extrapolation) in five CFs; (ii) fast, nearly sonic shear is resolved \cite{Keshet10} in $\sim$half of our CFs; and (iii) we find no correlation between $\xi$ and the data resolution.
Moreover, an unresolved layer yields $(1-\xi)\lesssim$ a few percent: for a layer width $w$ and a characteristic Mach number $\mach$, the normalized pressure jump induced is $P_{th}^{-1}\Delta P_{th}\approx \Gamma\mach^2 w/\rCF$, where $\Gamma=5/3$ is the adiabatic index; this holds for typical $n\propto r^{-1}$ and $\mach\propto r$ profiles beneath a concentric CF. With present resolutions, $w\lesssim 0.2r$. Even if $\mach\sim1$ flows were not already resolved, $\mach\lesssim 1$ implies that $P_{th}^{-1}\Delta P_{th}\lesssim$ a few $\%$.

To homogeneously test the deprojections, we repeat the surface brightness (SB) deprojection of the 7 CFs with published data and $\bar{\xi}=0.89\pm0.05$, finding consistent results with $\bar{\xi}=0.86\pm0.05$.
To test the $T$ deprojection, we combine deprojected $n$ with \emph{projected} $T$ profiles (available for three CFs \cite{Tanaka06,randall10}). As expected, $\xi$ slightly increases (by $0.04,0.1,0.16$), but is still $<1$.
Deprojection errors must be much smaller than this $<20\%$ effect, so any (smaller yet) offset in $\ave{\xi}$ is negligible.
Furthermore, we bin and deproject CF sectors in three cases with projected $T$ data \cite{Sanders09, sanders09a2204} in the method of \cite{MazzottaDeprojection}; no significant changes are found.

To test the extrapolation to $\rCF$, we examine three alternatives
(linear, logarithmic, and exponential) to the $T$ power-law fit.
All three give $\ave{\xi}=0.82\till 0.83$, consistent with Eq.~(\ref{eq:mean xi}), with insignificant changes in the control sample.
The $n$ profile is resolved and well fit by a broken power-law \cite[\eg][]{Markevitch00,Tanaka06}.
Finally, instead of two points on each side of the CF, we use all points within a factor $2$ from $\rCF$ (unless crossing another CF).
All $\xi$ values change within their error bars, resulting in $\ave{\xi}=0.85\pm0.04$.

We conclude that there are small but significant drops in $P_{th}$ below all nonmerger CFs, given by Eq.~(\ref{eq:mean xi}). Systematic effects enhance the dispersion among CFs, but do not significantly offset the average, $\ave{\xi}$. In fact, the deprojection errors needed to mimic our results greatly exceed the conservative estimates used in the analyses of CFs \cite{vikhlinin01,lal13,gastaldello13}, shocks \cite{GrahamEtAl08, blanton11,russell10,forman07}, sound waves, and physical mechanisms such as electron heating \cite{markevitch05, russell12}.

\emph{The missing pressure is nonthermal.---}
Observed CF pressure discontinuities are sometimes attributed to an unresolved stagnation region due to radial CF motion \cite{vikhlinin01}. Here, this would imply an inward acceleration of the CF, or outward ram pressure of the gas below the CF \cite{markevitch07}. However, a $10-20\%$ pressure jump would correspond \cite{vikhlinin01} to $\mach\simeq0.3-0.5$ \emph{radial} motion: unrealistic in an otherwise relaxed core. Moreover, core CFs form extended three-dimensional surfaces spanning much of the core, so the implied coherent radial motion is implausible.

As $P_{tot}$ is continuous across a steady CF, the missing pressure is naturally attributed to a nonthermal component. The pressure drop thus imposes a lower limit on the nonthermal pressure $P_{nt}$ just below the CF (an additional, smooth $P_{nt}$ component may exist). This depends on how the $\eta_{nt}(r) \equiv P_{nt}/\Ptot$ profile maps onto bins of radii $\{r_i\}$. An extended nonthermal layer corresponds to $\eta_{nt}(r<\rCF)\simeq 1-\xi$, whereas a narrow layer confined to the $i=-1$ temperature bin just below the CF gives
\begin{equation}\label{eq:eta-xi}
\eta_{nt,i=-1} \simeq 1-\xi^{\log\left({r_{-1}}/{r_{-2}}\right)/\log\left({r_{cf}}/{r_{-2}}\right)} \simeq 1-\xi^{{2}/{3}}
\end{equation}
in that bin, where $r_{-2}<r_{-1}$ are the mean radii in the two bins below the CF, and the last result holds for logarithmically spaced bins.
Averaging as in Eq.~(\ref{eq:mean xi}) yields $\ave{\eta}_{nt}(r<\rCF)=0.18\pm0.04$ or $\ave{\eta}_{nt,-1}=0.12\pm0.02$.

There is prior evidence for nonthermal pressure in clusters.
Comparing optical and X-ray data gives $\eta_{nt}\lesssim 10\%$ in the cores of Virgo and Fornax \cite{churazov08}. Comparing X-ray and weak lensing data yields $\eta_{nt}\sim30\%$ in MS2137 \cite{chiu12}.
However, these estimates assume hydrostatic equilibrium, and are not specific to CF regions.

\emph{The $P_{nt}$ distribution.---}
In 4 clusters, the deprojected $P_{th}$ profiles in CF sectors and azimuthally averaged can be compared (Figure \ref{fig:profiles}).
In Virgo and possibly 2A0335, $P_{th}$ above the CF is comparable to the mean pressure at that radius, whereas $P_{th}$ below the CF is smaller than the mean. This deviation is confined to $\sim[0.7,1]\rCF$, roughly corresponding to the shear layer \cite{Keshet10, Keshet12}, consistent with shear magnetization.

A2204 and possibly A496 suggest an opposite trend; $P_{th}$ \emph{exceeds} the azimuthal mean above the CF around $[0.9,1.5]\rCF$.
More accurate profiles are needed to resolve the underlying structure: the azimuthal means are biased by the normalization and by the presence of CFs, and the $P_{th}$ deviations are not highly significant.

\emph{The nonthermal pressure is mostly magnetic.---}
CF phenomenology strongly constrains the nature of the nonthermal pressure, as it is \emph{(i)} robustly found across different clusters; \emph{(ii)} seen at various radii within the core; \emph{(iii)} found near CFs; and \emph{(iv)} confined below the sharp CF transition. Shear-generated magnetic fields in the shear layers beneath CFs \cite{Keshet10} naturally fulfill all these conditions, as demonstrated numerically in Ref. \cite{zuhone11}. Three alternative explanations are examined and disfavored below: high-energy particles, small-scale turbulence (microturbulence), and a multiphase plasma.

We conclude that the most natural, and the only self-consistent, interpretation of the nonthermal pressure is at least predominantly magnetic.
As $\grad\cdot \bm{B}=0$, the discontinuity must be in the tangential field, in accordance with shear magnetization \cite{Keshet10} (this is also the field orientation driven by the heat-flux buoyancy instability \cite{parrish08}). Indeed, $\eta_{B}\gtrsim 0.1$ can explain the stability of CFs \cite{Keshet10,vikhlinin02}, the sharpness of the discontinuity, and the CF-minihalo connection.
Our most significantly nonzero field estimates, assuming $\eta_B=1-\xi^{2/3}$, are given in Table \ref{table:magnetic fields}.

\begin{table}[h]
\caption{\label{table:magnetic fields} Significant magnetization just below CFs (Eq.~\ref{eq:eta-xi}).}
\begin{ruledtabular}
\begin{tabular}{lcccc}
   Cluster & $\rCF$ $[\mbox{kpc}]$ & $\xi$ & $\eta_B$ & $B$ $[\mu$G$]$  \\
  \hline
  A133 & 19 & $0.69\pm0.08$ & $0.19\pm0.06$ & $34.0^{+6.3}_{-6.3}$ \\
  2A0335 & 29 & $0.78\pm0.09$ & $0.16\pm0.07$ & $25.1^{+6.3}_{-6.7}$ \\
  A496 & 159 & $0.75\pm0.13$ & $0.18\pm0.10$ & $14.4^{+4.9}_{-5.1}$ \\
  2A0335 & 62 & $0.73\pm0.17$ & $0.20\pm0.13$ & $21.8^{+9.2}_{-9.5}$ \\
  A1664 & 78 & $0.90\pm0.09$ & $0.07\pm0.06$ & $17.4^{+7.4}_{-12.2}$ \\
\end{tabular}
\end{ruledtabular}
\end{table}

\emph{Discussion.---} By analyzing all 17 deprojected core CF profiles from the literature, including 3 tentative new CFs (Figure \ref{fig:profiles}), we discover small but significant (Eq.~\ref{eq:mean xi}) drops below all nonmerger CFs (Figure \ref{fig:ratios}). Control samples (Figure \ref{fig:control profiles}) show no such trend; systematic effects are ruled out.
The most natural interpretation is an enhanced nonthermal pressure, predominantly magnetic, confined beneath the CF. The nonthermal fraction reaches $\eta_{nt}\simeq 10-20\%$ just below the CF. Such strong magnetization suffices to stabilize the CF, suppress transport across it, and explain the CF-minihalo coincidence.

The locally strong magnetization, its sharp rise below the CF, and its coincidence with the shear layer, suggest saturated magnetic shear-amplification. Note that magnetic saturation at similar, $\eta_B \sim 0.1$ levels, is sometimes inferred from observations of radio-bright regions in the ICM, in radio halos and relics \cite{keshet10relics}.
The typical amplitudes of the magnetic field, demonstrated in Table \ref{table:magnetic fields}, broadly agree with (the highly uncertain) Faraday rotation measurements \cite{clarke04} and minihalo estimates \cite{keshetloeb10}.

Although the CF narrowness and stability require the nonthermal pressure to be predominantly magnetic, $\lesssim 1/2$ of the localized $P_{nt}$ could be microturbulent.
However, upper limits on the projected velocity dispersion at the level of $5\till20\%$ were imposed in several clusters \cite{churazov08, sanders10turbulence, sanders11turbulence, bulbul12}.
Moreover, $\eta_B\gtrsim0.1$ would stabilize KHI, preventing the growth of turbulence in the first place.

Deviations from a single-phase gas can produce erroneous pressure profiles, distorting our analysis (\eg \cite{Tanaka06}). Such effects are unlikely to be significant, as this would require similar errors over a substantial range of temperatures and densities. Moreover, \emph{(i)} CC X-ray observations are consistent at any given position with a single phase \cite{molendi01};  \emph{(ii)} the CF pressure jump may appear \emph{larger} (by $\sim10\%$) in a multiphase model \cite{ghizzardi12}; and  \emph{(iii)} different multiphase plasmas on each side of the CF would still require strong magnetization, to stabilize the CF and isolate the plasma.

A diffuse high-energy component cannot explain our results. Even in the centers of CC \citep{keshetloeb10} and merger \citep{Kushnir2009,keshet10relics} clusters, its pressure fraction $\sim 10^{-3}$ is dynamically negligible. Shear acceleration \cite{rieger06} is inefficient in the ICM. AGN output is strongly diminished for $\rCF\gtrsim$ few $10\kpc$.
AGN bubbles could significantly contribute to $P_{nt}$; here too, the associated magnetic component must be large.

\acknowledgements
We thank the anonymous referees for helpful comments. This research has received funding from the European Union Seventh Framework Programme (FP7/2007-2013) under Grant Agreement n\textordmasculine ~293975, and from IAEC-UPBC Joint Research Foundation Grant 257/14.


\begin{thebibliography}{60}%
\makeatletter
\providecommand \@ifxundefined [1]{%
 \@ifx{#1\undefined}
}%
\providecommand \@ifnum [1]{%
 \ifnum #1\expandafter \@firstoftwo
 \else \expandafter \@secondoftwo
 \fi
}%
\providecommand \@ifx [1]{%
 \ifx #1\expandafter \@firstoftwo
 \else \expandafter \@secondoftwo
 \fi
}%
\providecommand \natexlab [1]{#1}%
\providecommand \enquote  [1]{``#1''}%
\providecommand \bibnamefont  [1]{#1}%
\providecommand \bibfnamefont [1]{#1}%
\providecommand \citenamefont [1]{#1}%
\providecommand \href@noop [0]{\@secondoftwo}%
\providecommand \href [0]{\begingroup \@sanitize@url \@href}%
\providecommand \@href[1]{\@@startlink{#1}\@@href}%
\providecommand \@@href[1]{\endgroup#1\@@endlink}%
\providecommand \@sanitize@url [0]{\catcode `\\12\catcode `\$12\catcode
  `\&12\catcode `\#12\catcode `\^12\catcode `\_12\catcode `\%12\relax}%
\providecommand \@@startlink[1]{}%
\providecommand \@@endlink[0]{}%
\providecommand \url  [0]{\begingroup\@sanitize@url \@url }%
\providecommand \@url [1]{\endgroup\@href {#1}{\urlprefix }}%
\providecommand \urlprefix  [0]{URL }%
\providecommand \Eprint [0]{\href }%
\providecommand \doibase [0]{http://dx.doi.org/}%
\providecommand \selectlanguage [0]{\@gobble}%
\providecommand \bibinfo  [0]{\@secondoftwo}%
\providecommand \bibfield  [0]{\@secondoftwo}%
\providecommand \translation [1]{[#1]}%
\providecommand \BibitemOpen [0]{}%
\providecommand \bibitemStop [0]{}%
\providecommand \bibitemNoStop [0]{.\EOS\space}%
\providecommand \EOS [0]{\spacefactor3000\relax}%
\providecommand \BibitemShut  [1]{\csname bibitem#1\endcsname}%
\let\auto@bib@innerbib\@empty
\bibitem [{\citenamefont {{Markevitch}}\ and\ \citenamefont
  {{Vikhlinin}}(2007)}]{markevitch07}%
  \BibitemOpen
  \bibfield  {author} {\bibinfo {author} {\bibfnamefont {M.}~\bibnamefont
  {{Markevitch}}}\ and\ \bibinfo {author} {\bibfnamefont {A.}~\bibnamefont
  {{Vikhlinin}}},\ }\href {\doibase 10.1016/j.physrep.2007.01.001} {\bibfield
  {journal} {\bibinfo  {journal} {\physrep}\ }\textbf {\bibinfo {volume}
  {443}},\ \bibinfo {pages} {1} (\bibinfo {year} {2007})},\ \Eprint
  {http://arxiv.org/abs/arXiv:astro-ph/0701821} {arXiv:astro-ph/0701821}
  \BibitemShut {NoStop}%
\bibitem [{\citenamefont {{Markevitch}}\ \emph {et~al.}(2003)\citenamefont
  {{Markevitch}}, \citenamefont {{Vikhlinin}},\ and\ \citenamefont
  {{Forman}}}]{Markevitch03}%
  \BibitemOpen
  \bibfield  {author} {\bibinfo {author} {\bibfnamefont {M.}~\bibnamefont
  {{Markevitch}}}, \bibinfo {author} {\bibfnamefont {A.}~\bibnamefont
  {{Vikhlinin}}}, \ and\ \bibinfo {author} {\bibfnamefont {W.~R.}\ \bibnamefont
  {{Forman}}},\ }in\ \href@noop {} {\emph {\bibinfo {booktitle} {Astronomical
  Society of the Pacific Conference Series}}},\ Vol.\ \bibinfo {volume} {301},\
  \bibinfo {editor} {edited by\ \bibinfo {editor} {\bibfnamefont
  {S.}~\bibnamefont {{Bowyer}}}\ and\ \bibinfo {editor} {\bibfnamefont {C.-Y.}\
  \bibnamefont {{Hwang}}}}\ (\bibinfo {year} {2003})\ p.~\bibinfo {pages}
  {37},\ \Eprint {http://arxiv.org/abs/arXiv:astro-ph/0208208}
  {arXiv:astro-ph/0208208} \BibitemShut {NoStop}%
\bibitem [{\citenamefont {{Owers}}\ \emph {et~al.}(2009)\citenamefont
  {{Owers}}, \citenamefont {{Nulsen}}, \citenamefont {{Couch}},\ and\
  \citenamefont {{Markevitch}}}]{owers09}%
  \BibitemOpen
  \bibfield  {author} {\bibinfo {author} {\bibfnamefont {M.~S.}\ \bibnamefont
  {{Owers}}}, \bibinfo {author} {\bibfnamefont {P.~E.~J.}\ \bibnamefont
  {{Nulsen}}}, \bibinfo {author} {\bibfnamefont {W.~J.}\ \bibnamefont
  {{Couch}}}, \ and\ \bibinfo {author} {\bibfnamefont {M.}~\bibnamefont
  {{Markevitch}}},\ }\href {\doibase 10.1088/0004-637X/704/2/1349} {\bibfield
  {journal} {\bibinfo  {journal} {\apj}\ }\textbf {\bibinfo {volume} {704}},\
  \bibinfo {pages} {1349} (\bibinfo {year} {2009})},\ \Eprint
  {http://arxiv.org/abs/0909.2645} {arXiv:0909.2645 [astro-ph.CO]} \BibitemShut
  {NoStop}%
\bibitem [{\citenamefont {{Ascasibar}}\ and\ \citenamefont
  {{Markevitch}}(2006)}]{ascasibar06}%
  \BibitemOpen
  \bibfield  {author} {\bibinfo {author} {\bibfnamefont {Y.}~\bibnamefont
  {{Ascasibar}}}\ and\ \bibinfo {author} {\bibfnamefont {M.}~\bibnamefont
  {{Markevitch}}},\ }\href {\doibase 10.1086/506508} {\bibfield  {journal}
  {\bibinfo  {journal} {\apj}\ }\textbf {\bibinfo {volume} {650}},\ \bibinfo
  {pages} {102} (\bibinfo {year} {2006})},\ \Eprint
  {http://arxiv.org/abs/arXiv:astro-ph/0603246} {arXiv:astro-ph/0603246}
  \BibitemShut {NoStop}%
\bibitem [{\citenamefont {{Keshet}}(2012)}]{Keshet12}%
  \BibitemOpen
  \bibfield  {author} {\bibinfo {author} {\bibfnamefont {U.}~\bibnamefont
  {{Keshet}}},\ }\href {\doibase 10.1088/0004-637X/753/2/120} {\bibfield
  {journal} {\bibinfo  {journal} {\apj}\ }\textbf {\bibinfo {volume} {753}},\
  \bibinfo {eid} {120} (\bibinfo {year} {2012})},\ \Eprint
  {http://arxiv.org/abs/1111.2337} {arXiv:1111.2337 [astro-ph.CO]} \BibitemShut
  {NoStop}%
\bibitem [{\citenamefont {{Markevitch}}\ \emph {et~al.}(2001)\citenamefont
  {{Markevitch}}, \citenamefont {{Vikhlinin}},\ and\ \citenamefont
  {{Mazzotta}}}]{Markevitch01}%
  \BibitemOpen
  \bibfield  {author} {\bibinfo {author} {\bibfnamefont {M.}~\bibnamefont
  {{Markevitch}}}, \bibinfo {author} {\bibfnamefont {A.}~\bibnamefont
  {{Vikhlinin}}}, \ and\ \bibinfo {author} {\bibfnamefont {P.}~\bibnamefont
  {{Mazzotta}}},\ }\href {\doibase 10.1086/337973} {\bibfield  {journal}
  {\bibinfo  {journal} {\apjl}\ }\textbf {\bibinfo {volume} {562}},\ \bibinfo
  {pages} {L153} (\bibinfo {year} {2001})},\ \Eprint
  {http://arxiv.org/abs/arXiv:astro-ph/0108520} {arXiv:astro-ph/0108520}
  \BibitemShut {NoStop}%
\bibitem [{\citenamefont {{Tittley}}\ and\ \citenamefont
  {{Henriksen}}(2005)}]{Tittley05}%
  \BibitemOpen
  \bibfield  {author} {\bibinfo {author} {\bibfnamefont {E.~R.}\ \bibnamefont
  {{Tittley}}}\ and\ \bibinfo {author} {\bibfnamefont {M.}~\bibnamefont
  {{Henriksen}}},\ }\href {\doibase 10.1086/425952} {\bibfield  {journal}
  {\bibinfo  {journal} {\apj}\ }\textbf {\bibinfo {volume} {618}},\ \bibinfo
  {pages} {227} (\bibinfo {year} {2005})},\ \Eprint
  {http://arxiv.org/abs/arXiv:astro-ph/0409177} {arXiv:astro-ph/0409177}
  \BibitemShut {NoStop}%
\bibitem [{\citenamefont {{Churazov}}\ \emph {et~al.}(2003)\citenamefont
  {{Churazov}}, \citenamefont {{Forman}}, \citenamefont {{Jones}},\ and\
  \citenamefont {{B{\"o}hringer}}}]{Churazov03}%
  \BibitemOpen
  \bibfield  {author} {\bibinfo {author} {\bibfnamefont {E.}~\bibnamefont
  {{Churazov}}}, \bibinfo {author} {\bibfnamefont {W.}~\bibnamefont
  {{Forman}}}, \bibinfo {author} {\bibfnamefont {C.}~\bibnamefont {{Jones}}}, \
  and\ \bibinfo {author} {\bibfnamefont {H.}~\bibnamefont {{B{\"o}hringer}}},\
  }\href {\doibase 10.1086/374923} {\bibfield  {journal} {\bibinfo  {journal}
  {\apj}\ }\textbf {\bibinfo {volume} {590}},\ \bibinfo {pages} {225} (\bibinfo
  {year} {2003})},\ \Eprint {http://arxiv.org/abs/arXiv:astro-ph/0301482}
  {arXiv:astro-ph/0301482} \BibitemShut {NoStop}%
\bibitem [{\citenamefont {{Fujita}}\ \emph {et~al.}(2004)\citenamefont
  {{Fujita}}, \citenamefont {{Matsumoto}},\ and\ \citenamefont
  {{Wada}}}]{fujita04}%
  \BibitemOpen
  \bibfield  {author} {\bibinfo {author} {\bibfnamefont {Y.}~\bibnamefont
  {{Fujita}}}, \bibinfo {author} {\bibfnamefont {T.}~\bibnamefont
  {{Matsumoto}}}, \ and\ \bibinfo {author} {\bibfnamefont {K.}~\bibnamefont
  {{Wada}}},\ }\href {\doibase 10.1086/424483} {\bibfield  {journal} {\bibinfo
  {journal} {\apjl}\ }\textbf {\bibinfo {volume} {612}},\ \bibinfo {pages} {L9}
  (\bibinfo {year} {2004})},\ \Eprint
  {http://arxiv.org/abs/arXiv:astro-ph/0407368} {arXiv:astro-ph/0407368}
  \BibitemShut {NoStop}%
\bibitem [{\citenamefont {{Keshet}}\ \emph {et~al.}(2010)\citenamefont
  {{Keshet}}, \citenamefont {{Markevitch}}, \citenamefont {{Birnboim}},\ and\
  \citenamefont {{Loeb}}}]{Keshet10}%
  \BibitemOpen
  \bibfield  {author} {\bibinfo {author} {\bibfnamefont {U.}~\bibnamefont
  {{Keshet}}}, \bibinfo {author} {\bibfnamefont {M.}~\bibnamefont
  {{Markevitch}}}, \bibinfo {author} {\bibfnamefont {Y.}~\bibnamefont
  {{Birnboim}}}, \ and\ \bibinfo {author} {\bibfnamefont {A.}~\bibnamefont
  {{Loeb}}},\ }\href {\doibase 10.1088/2041-8205/719/1/L74} {\bibfield
  {journal} {\bibinfo  {journal} {\apjl}\ }\textbf {\bibinfo {volume} {719}},\
  \bibinfo {pages} {L74} (\bibinfo {year} {2010})},\ \Eprint
  {http://arxiv.org/abs/0912.3526} {arXiv:0912.3526 [astro-ph.CO]} \BibitemShut
  {NoStop}%
\bibitem [{\citenamefont {{ZuHone}}\ \emph {et~al.}(2011)\citenamefont
  {{ZuHone}}, \citenamefont {{Markevitch}},\ and\ \citenamefont
  {{Lee}}}]{zuhone11}%
  \BibitemOpen
  \bibfield  {author} {\bibinfo {author} {\bibfnamefont {J.~A.}\ \bibnamefont
  {{ZuHone}}}, \bibinfo {author} {\bibfnamefont {M.}~\bibnamefont
  {{Markevitch}}}, \ and\ \bibinfo {author} {\bibfnamefont {D.}~\bibnamefont
  {{Lee}}},\ }\href {\doibase 10.1088/0004-637X/743/1/16} {\bibfield  {journal}
  {\bibinfo  {journal} {\apj}\ }\textbf {\bibinfo {volume} {743}},\ \bibinfo
  {eid} {16} (\bibinfo {year} {2011})},\ \Eprint
  {http://arxiv.org/abs/1108.4427} {arXiv:1108.4427 [astro-ph.CO]} \BibitemShut
  {NoStop}%
\bibitem [{\citenamefont {{Vikhlinin}}\ \emph {et~al.}(2001)\citenamefont
  {{Vikhlinin}}, \citenamefont {{Markevitch}},\ and\ \citenamefont
  {{Murray}}}]{vikhlinin01}%
  \BibitemOpen
  \bibfield  {author} {\bibinfo {author} {\bibfnamefont {A.}~\bibnamefont
  {{Vikhlinin}}}, \bibinfo {author} {\bibfnamefont {M.}~\bibnamefont
  {{Markevitch}}}, \ and\ \bibinfo {author} {\bibfnamefont {S.~S.}\
  \bibnamefont {{Murray}}},\ }\href {\doibase 10.1086/320078} {\bibfield
  {journal} {\bibinfo  {journal} {\apj}\ }\textbf {\bibinfo {volume} {551}},\
  \bibinfo {pages} {160} (\bibinfo {year} {2001})},\ \Eprint
  {http://arxiv.org/abs/arXiv:astro-ph/0008496} {arXiv:astro-ph/0008496}
  \BibitemShut {NoStop}%
\bibitem [{\citenamefont {{Ettori}}\ and\ \citenamefont
  {{Fabian}}(2000)}]{ettori00}%
  \BibitemOpen
  \bibfield  {author} {\bibinfo {author} {\bibfnamefont {S.}~\bibnamefont
  {{Ettori}}}\ and\ \bibinfo {author} {\bibfnamefont {A.~C.}\ \bibnamefont
  {{Fabian}}},\ }\href {\doibase 10.1046/j.1365-8711.2000.03899.x} {\bibfield
  {journal} {\bibinfo  {journal} {\mnras}\ }\textbf {\bibinfo {volume} {317}},\
  \bibinfo {pages} {L57} (\bibinfo {year} {2000})},\ \Eprint
  {http://arxiv.org/abs/arXiv:astro-ph/0007397} {arXiv:astro-ph/0007397}
  \BibitemShut {NoStop}%
\bibitem [{\citenamefont {{Mazzotta}}\ and\ \citenamefont
  {{Giacintucci}}(2008)}]{mazzotta08}%
  \BibitemOpen
  \bibfield  {author} {\bibinfo {author} {\bibfnamefont {P.}~\bibnamefont
  {{Mazzotta}}}\ and\ \bibinfo {author} {\bibfnamefont {S.}~\bibnamefont
  {{Giacintucci}}},\ }\href {\doibase 10.1086/529433} {\bibfield  {journal}
  {\bibinfo  {journal} {\apjl}\ }\textbf {\bibinfo {volume} {675}},\ \bibinfo
  {pages} {L9} (\bibinfo {year} {2008})},\ \Eprint
  {http://arxiv.org/abs/0801.1905} {arXiv:0801.1905} \BibitemShut {NoStop}%
\bibitem [{\citenamefont {{Giacintucci}}\ \emph {et~al.}(2011)\citenamefont
  {{Giacintucci}}, \citenamefont {{Markevitch}}, \citenamefont {{Brunetti}},
  \citenamefont {{Cassano}},\ and\ \citenamefont {{Venturi}}}]{giacintucci11}%
  \BibitemOpen
  \bibfield  {author} {\bibinfo {author} {\bibfnamefont {S.}~\bibnamefont
  {{Giacintucci}}}, \bibinfo {author} {\bibfnamefont {M.}~\bibnamefont
  {{Markevitch}}}, \bibinfo {author} {\bibfnamefont {G.}~\bibnamefont
  {{Brunetti}}}, \bibinfo {author} {\bibfnamefont {R.}~\bibnamefont
  {{Cassano}}}, \ and\ \bibinfo {author} {\bibfnamefont {T.}~\bibnamefont
  {{Venturi}}},\ }\href {\doibase 10.1051/0004-6361/201015882} {\bibfield
  {journal} {\bibinfo  {journal} {\aap}\ }\textbf {\bibinfo {volume} {525}},\
  \bibinfo {eid} {L10} (\bibinfo {year} {2011})},\ \Eprint
  {http://arxiv.org/abs/1011.3141} {arXiv:1011.3141 [astro-ph.HE]} \BibitemShut
  {NoStop}%
\bibitem [{\citenamefont {{ZuHone}}\ \emph {et~al.}(2013)\citenamefont
  {{ZuHone}}, \citenamefont {{Markevitch}}, \citenamefont {{Brunetti}},\ and\
  \citenamefont {{Giacintucci}}}]{zuhone12}%
  \BibitemOpen
  \bibfield  {author} {\bibinfo {author} {\bibfnamefont {J.~A.}\ \bibnamefont
  {{ZuHone}}}, \bibinfo {author} {\bibfnamefont {M.}~\bibnamefont
  {{Markevitch}}}, \bibinfo {author} {\bibfnamefont {G.}~\bibnamefont
  {{Brunetti}}}, \ and\ \bibinfo {author} {\bibfnamefont {S.}~\bibnamefont
  {{Giacintucci}}},\ }\href {\doibase 10.1088/0004-637X/762/2/78} {\bibfield
  {journal} {\bibinfo  {journal} {\apj}\ }\textbf {\bibinfo {volume} {762}},\
  \bibinfo {eid} {78} (\bibinfo {year} {2013})},\ \Eprint
  {http://arxiv.org/abs/1203.2994} {arXiv:1203.2994 [astro-ph.CO]} \BibitemShut
  {NoStop}%
\bibitem [{\citenamefont {{Keshet}}\ and\ \citenamefont
  {{Loeb}}(2010)}]{keshetloeb10}%
  \BibitemOpen
  \bibfield  {author} {\bibinfo {author} {\bibfnamefont {U.}~\bibnamefont
  {{Keshet}}}\ and\ \bibinfo {author} {\bibfnamefont {A.}~\bibnamefont
  {{Loeb}}},\ }\href {\doibase 10.1088/0004-637X/722/1/737} {\bibfield
  {journal} {\bibinfo  {journal} {\apj}\ }\textbf {\bibinfo {volume} {722}},\
  \bibinfo {pages} {737} (\bibinfo {year} {2010})},\ \Eprint
  {http://arxiv.org/abs/1003.1133} {arXiv:1003.1133 [astro-ph.CO]} \BibitemShut
  {NoStop}%
\bibitem [{\citenamefont {{Carilli}}\ and\ \citenamefont
  {{Taylor}}(2002)}]{carilli02}%
  \BibitemOpen
  \bibfield  {author} {\bibinfo {author} {\bibfnamefont {C.~L.}\ \bibnamefont
  {{Carilli}}}\ and\ \bibinfo {author} {\bibfnamefont {G.~B.}\ \bibnamefont
  {{Taylor}}},\ }\href {\doibase 10.1146/annurev.astro.40.060401.093852}
  {\bibfield  {journal} {\bibinfo  {journal} {\araa}\ }\textbf {\bibinfo
  {volume} {40}},\ \bibinfo {pages} {319} (\bibinfo {year} {2002})},\ \Eprint
  {http://arxiv.org/abs/arXiv:astro-ph/0110655} {arXiv:astro-ph/0110655}
  \BibitemShut {NoStop}%
\bibitem [{\citenamefont {{Clarke}}(2004)}]{clarke04}%
  \BibitemOpen
  \bibfield  {author} {\bibinfo {author} {\bibfnamefont {T.~E.}\ \bibnamefont
  {{Clarke}}},\ }\href@noop {} {\bibfield  {journal} {\bibinfo  {journal}
  {Journal of Korean Astronomical Society}\ }\textbf {\bibinfo {volume} {37}},\
  \bibinfo {pages} {337} (\bibinfo {year} {2004})},\ \Eprint
  {http://arxiv.org/abs/arXiv:astro-ph/0412268} {arXiv:astro-ph/0412268}
  \BibitemShut {NoStop}%
\bibitem [{\citenamefont {{Govoni}}\ \emph {et~al.}(2010)\citenamefont
  {{Govoni}}, \citenamefont {{Dolag}}, \citenamefont {{Murgia}}, \citenamefont
  {{Feretti}}, \citenamefont {{Schindler}}, \citenamefont {{Giovannini}},
  \citenamefont {{Boschin}}, \citenamefont {{Vacca}},\ and\ \citenamefont
  {{Bonafede}}}]{govoni10}%
  \BibitemOpen
  \bibfield  {author} {\bibinfo {author} {\bibfnamefont {F.}~\bibnamefont
  {{Govoni}}}, \bibinfo {author} {\bibfnamefont {K.}~\bibnamefont {{Dolag}}},
  \bibinfo {author} {\bibfnamefont {M.}~\bibnamefont {{Murgia}}}, \bibinfo
  {author} {\bibfnamefont {L.}~\bibnamefont {{Feretti}}}, \bibinfo {author}
  {\bibfnamefont {S.}~\bibnamefont {{Schindler}}}, \bibinfo {author}
  {\bibfnamefont {G.}~\bibnamefont {{Giovannini}}}, \bibinfo {author}
  {\bibfnamefont {W.}~\bibnamefont {{Boschin}}}, \bibinfo {author}
  {\bibfnamefont {V.}~\bibnamefont {{Vacca}}}, \ and\ \bibinfo {author}
  {\bibfnamefont {A.}~\bibnamefont {{Bonafede}}},\ }\href {\doibase
  10.1051/0004-6361/200913665} {\bibfield  {journal} {\bibinfo  {journal}
  {\aap}\ }\textbf {\bibinfo {volume} {522}},\ \bibinfo {eid} {A105} (\bibinfo
  {year} {2010})},\ \Eprint {http://arxiv.org/abs/1007.5207} {arXiv:1007.5207
  [astro-ph.CO]} \BibitemShut {NoStop}%
\bibitem [{\citenamefont {{Bonafede}}\ \emph {et~al.}(2011)\citenamefont
  {{Bonafede}}, \citenamefont {{Govoni}}, \citenamefont {{Feretti}},
  \citenamefont {{Murgia}}, \citenamefont {{Giovannini}},\ and\ \citenamefont
  {{Br{\"u}ggen}}}]{Bonafede11}%
  \BibitemOpen
  \bibfield  {author} {\bibinfo {author} {\bibfnamefont {A.}~\bibnamefont
  {{Bonafede}}}, \bibinfo {author} {\bibfnamefont {F.}~\bibnamefont
  {{Govoni}}}, \bibinfo {author} {\bibfnamefont {L.}~\bibnamefont {{Feretti}}},
  \bibinfo {author} {\bibfnamefont {M.}~\bibnamefont {{Murgia}}}, \bibinfo
  {author} {\bibfnamefont {G.}~\bibnamefont {{Giovannini}}}, \ and\ \bibinfo
  {author} {\bibfnamefont {M.}~\bibnamefont {{Br{\"u}ggen}}},\ }\href {\doibase
  10.1051/0004-6361/201016298} {\bibfield  {journal} {\bibinfo  {journal}
  {\aap}\ }\textbf {\bibinfo {volume} {530}},\ \bibinfo {eid} {A24} (\bibinfo
  {year} {2011})}\BibitemShut {NoStop}%
\bibitem [{\citenamefont {{Tamura}}\ \emph {et~al.}(2003)\citenamefont
  {{Tamura}}, \citenamefont {{Kaastra}}, \citenamefont {{Bleeker}},\ and\
  \citenamefont {{Peterson}}}]{tamura03}%
  \BibitemOpen
  \bibfield  {author} {\bibinfo {author} {\bibfnamefont {T.}~\bibnamefont
  {{Tamura}}}, \bibinfo {author} {\bibfnamefont {J.~S.}\ \bibnamefont
  {{Kaastra}}}, \bibinfo {author} {\bibfnamefont {J.~A.~M.}\ \bibnamefont
  {{Bleeker}}}, \ and\ \bibinfo {author} {\bibfnamefont {J.~R.}\ \bibnamefont
  {{Peterson}}},\ }in\ \href@noop {} {\emph {\bibinfo {booktitle} {Workshop on
  Galaxies and Clusters of Galaxies}}}\ (\bibinfo {year} {2003})\ p.~\bibinfo
  {pages} {65}\BibitemShut {NoStop}%
\bibitem [{\citenamefont {{Roediger}}\ \emph {et~al.}(2011)\citenamefont
  {{Roediger}}, \citenamefont {{Br{\"u}ggen}}, \citenamefont {{Simionescu}},
  \citenamefont {{B{\"o}hringer}}, \citenamefont {{Churazov}},\ and\
  \citenamefont {{Forman}}}]{roediger11}%
  \BibitemOpen
  \bibfield  {author} {\bibinfo {author} {\bibfnamefont {E.}~\bibnamefont
  {{Roediger}}}, \bibinfo {author} {\bibfnamefont {M.}~\bibnamefont
  {{Br{\"u}ggen}}}, \bibinfo {author} {\bibfnamefont {A.}~\bibnamefont
  {{Simionescu}}}, \bibinfo {author} {\bibfnamefont {H.}~\bibnamefont
  {{B{\"o}hringer}}}, \bibinfo {author} {\bibfnamefont {E.}~\bibnamefont
  {{Churazov}}}, \ and\ \bibinfo {author} {\bibfnamefont {W.~R.}\ \bibnamefont
  {{Forman}}},\ }\href {\doibase 10.1111/j.1365-2966.2011.18279.x} {\bibfield
  {journal} {\bibinfo  {journal} {\mnras}\ }\textbf {\bibinfo {volume} {413}},\
  \bibinfo {pages} {2057} (\bibinfo {year} {2011})},\ \Eprint
  {http://arxiv.org/abs/1007.4209} {arXiv:1007.4209 [astro-ph.CO]} \BibitemShut
  {NoStop}%
\bibitem [{\citenamefont {{Sanders}}\ \emph
  {et~al.}(2009{\natexlab{a}})\citenamefont {{Sanders}}, \citenamefont
  {{Fabian}},\ and\ \citenamefont {{Taylor}}}]{sanders09a2204}%
  \BibitemOpen
  \bibfield  {author} {\bibinfo {author} {\bibfnamefont {J.~S.}\ \bibnamefont
  {{Sanders}}}, \bibinfo {author} {\bibfnamefont {A.~C.}\ \bibnamefont
  {{Fabian}}}, \ and\ \bibinfo {author} {\bibfnamefont {G.~B.}\ \bibnamefont
  {{Taylor}}},\ }\href {\doibase 10.1111/j.1365-2966.2008.14207.x} {\bibfield
  {journal} {\bibinfo  {journal} {\mnras}\ }\textbf {\bibinfo {volume} {393}},\
  \bibinfo {pages} {71} (\bibinfo {year} {2009}{\natexlab{a}})},\ \Eprint
  {http://arxiv.org/abs/0811.0743} {arXiv:0811.0743} \BibitemShut {NoStop}%
\bibitem [{\citenamefont {{Sanders}}\ \emph {et~al.}(2005)\citenamefont
  {{Sanders}}, \citenamefont {{Fabian}},\ and\ \citenamefont
  {{Taylor}}}]{Sanders05}%
  \BibitemOpen
  \bibfield  {author} {\bibinfo {author} {\bibfnamefont {J.~S.}\ \bibnamefont
  {{Sanders}}}, \bibinfo {author} {\bibfnamefont {A.~C.}\ \bibnamefont
  {{Fabian}}}, \ and\ \bibinfo {author} {\bibfnamefont {G.~B.}\ \bibnamefont
  {{Taylor}}},\ }\href {\doibase 10.1111/j.1365-2966.2004.08526.x} {\bibfield
  {journal} {\bibinfo  {journal} {\mnras}\ }\textbf {\bibinfo {volume} {356}},\
  \bibinfo {pages} {1022} (\bibinfo {year} {2005})},\ \Eprint
  {http://arxiv.org/abs/arXiv:astro-ph/0406094} {arXiv:astro-ph/0406094}
  \BibitemShut {NoStop}%
\bibitem [{\citenamefont {{Tanaka}}\ \emph {et~al.}(2006)\citenamefont
  {{Tanaka}}, \citenamefont {{Kunieda}}, \citenamefont {{Hudaverdi}},
  \citenamefont {{Furuzawa}},\ and\ \citenamefont {{Tawara}}}]{Tanaka06}%
  \BibitemOpen
  \bibfield  {author} {\bibinfo {author} {\bibfnamefont {T.}~\bibnamefont
  {{Tanaka}}}, \bibinfo {author} {\bibfnamefont {H.}~\bibnamefont {{Kunieda}}},
  \bibinfo {author} {\bibfnamefont {M.}~\bibnamefont {{Hudaverdi}}}, \bibinfo
  {author} {\bibfnamefont {A.}~\bibnamefont {{Furuzawa}}}, \ and\ \bibinfo
  {author} {\bibfnamefont {Y.}~\bibnamefont {{Tawara}}},\ }\href@noop {}
  {\bibfield  {journal} {\bibinfo  {journal} {\pasj}\ }\textbf {\bibinfo
  {volume} {58}},\ \bibinfo {pages} {703} (\bibinfo {year} {2006})}\BibitemShut
  {NoStop}%
\bibitem [{\citenamefont {{Sanders}}\ \emph
  {et~al.}(2009{\natexlab{b}})\citenamefont {{Sanders}}, \citenamefont
  {{Fabian}},\ and\ \citenamefont {{Taylor}}}]{Sanders09}%
  \BibitemOpen
  \bibfield  {author} {\bibinfo {author} {\bibfnamefont {J.~S.}\ \bibnamefont
  {{Sanders}}}, \bibinfo {author} {\bibfnamefont {A.~C.}\ \bibnamefont
  {{Fabian}}}, \ and\ \bibinfo {author} {\bibfnamefont {G.~B.}\ \bibnamefont
  {{Taylor}}},\ }\href {\doibase 10.1111/j.1365-2966.2009.14892.x} {\bibfield
  {journal} {\bibinfo  {journal} {\mnras}\ }\textbf {\bibinfo {volume} {396}},\
  \bibinfo {pages} {1449} (\bibinfo {year} {2009}{\natexlab{b}})},\ \Eprint
  {http://arxiv.org/abs/0904.1374} {arXiv:0904.1374 [astro-ph.CO]} \BibitemShut
  {NoStop}%
\bibitem [{\citenamefont {{Urban}}\ \emph {et~al.}(2011)\citenamefont
  {{Urban}}, \citenamefont {{Werner}}, \citenamefont {{Simionescu}},
  \citenamefont {{Allen}},\ and\ \citenamefont {{B{\"o}hringer}}}]{urban11}%
  \BibitemOpen
  \bibfield  {author} {\bibinfo {author} {\bibfnamefont {O.}~\bibnamefont
  {{Urban}}}, \bibinfo {author} {\bibfnamefont {N.}~\bibnamefont {{Werner}}},
  \bibinfo {author} {\bibfnamefont {A.}~\bibnamefont {{Simionescu}}}, \bibinfo
  {author} {\bibfnamefont {S.~W.}\ \bibnamefont {{Allen}}}, \ and\ \bibinfo
  {author} {\bibfnamefont {H.}~\bibnamefont {{B{\"o}hringer}}},\ }\href
  {\doibase 10.1111/j.1365-2966.2011.18526.x} {\bibfield  {journal} {\bibinfo
  {journal} {\mnras}\ }\textbf {\bibinfo {volume} {414}},\ \bibinfo {pages}
  {2101} (\bibinfo {year} {2011})},\ \Eprint {http://arxiv.org/abs/1102.2430}
  {arXiv:1102.2430 [astro-ph.CO]} \BibitemShut {NoStop}%
\bibitem [{\citenamefont {{Kirkpatrick}}\ \emph {et~al.}(2009)\citenamefont
  {{Kirkpatrick}}, \citenamefont {{McNamara}}, \citenamefont {{Rafferty}},
  \citenamefont {{Nulsen}}, \citenamefont {{B{\^i}rzan}}, \citenamefont
  {{Kazemzadeh}}, \citenamefont {{Wise}}, \citenamefont {{Gitti}},\ and\
  \citenamefont {{Cavagnolo}}}]{kirkpatrick09}%
  \BibitemOpen
  \bibfield  {author} {\bibinfo {author} {\bibfnamefont {C.~C.}\ \bibnamefont
  {{Kirkpatrick}}}, \bibinfo {author} {\bibfnamefont {B.~R.}\ \bibnamefont
  {{McNamara}}}, \bibinfo {author} {\bibfnamefont {D.~A.}\ \bibnamefont
  {{Rafferty}}}, \bibinfo {author} {\bibfnamefont {P.~E.~J.}\ \bibnamefont
  {{Nulsen}}}, \bibinfo {author} {\bibfnamefont {L.}~\bibnamefont
  {{B{\^i}rzan}}}, \bibinfo {author} {\bibfnamefont {F.}~\bibnamefont
  {{Kazemzadeh}}}, \bibinfo {author} {\bibfnamefont {M.~W.}\ \bibnamefont
  {{Wise}}}, \bibinfo {author} {\bibfnamefont {M.}~\bibnamefont {{Gitti}}}, \
  and\ \bibinfo {author} {\bibfnamefont {K.~W.}\ \bibnamefont {{Cavagnolo}}},\
  }\href {\doibase 10.1088/0004-637X/697/1/867} {\bibfield  {journal} {\bibinfo
   {journal} {\apj}\ }\textbf {\bibinfo {volume} {697}},\ \bibinfo {pages}
  {867} (\bibinfo {year} {2009})},\ \Eprint {http://arxiv.org/abs/0903.1108}
  {arXiv:0903.1108 [astro-ph.GA]} \BibitemShut {NoStop}%
\bibitem [{\citenamefont {{de Plaa}}\ \emph {et~al.}(2010)\citenamefont {{de
  Plaa}}, \citenamefont {{Werner}}, \citenamefont {{Simionescu}}, \citenamefont
  {{Kaastra}}, \citenamefont {{Grange}},\ and\ \citenamefont
  {{Vink}}}]{dePlaa10}%
  \BibitemOpen
  \bibfield  {author} {\bibinfo {author} {\bibfnamefont {J.}~\bibnamefont {{de
  Plaa}}}, \bibinfo {author} {\bibfnamefont {N.}~\bibnamefont {{Werner}}},
  \bibinfo {author} {\bibfnamefont {A.}~\bibnamefont {{Simionescu}}}, \bibinfo
  {author} {\bibfnamefont {J.~S.}\ \bibnamefont {{Kaastra}}}, \bibinfo {author}
  {\bibfnamefont {Y.~G.}\ \bibnamefont {{Grange}}}, \ and\ \bibinfo {author}
  {\bibfnamefont {J.}~\bibnamefont {{Vink}}},\ }\href {\doibase
  10.1051/0004-6361/201015198} {\bibfield  {journal} {\bibinfo  {journal}
  {\aap}\ }\textbf {\bibinfo {volume} {523}},\ \bibinfo {eid} {A81} (\bibinfo
  {year} {2010})},\ \Eprint {http://arxiv.org/abs/1008.3109} {arXiv:1008.3109
  [astro-ph.CO]} \BibitemShut {NoStop}%
\bibitem [{\citenamefont {{Johnson}}\ \emph {et~al.}(2010)\citenamefont
  {{Johnson}}, \citenamefont {{Markevitch}}, \citenamefont {{Wegner}},
  \citenamefont {{Jones}},\ and\ \citenamefont {{Forman}}}]{Johnson10}%
  \BibitemOpen
  \bibfield  {author} {\bibinfo {author} {\bibfnamefont {R.~E.}\ \bibnamefont
  {{Johnson}}}, \bibinfo {author} {\bibfnamefont {M.}~\bibnamefont
  {{Markevitch}}}, \bibinfo {author} {\bibfnamefont {G.~A.}\ \bibnamefont
  {{Wegner}}}, \bibinfo {author} {\bibfnamefont {C.}~\bibnamefont {{Jones}}}, \
  and\ \bibinfo {author} {\bibfnamefont {W.~R.}\ \bibnamefont {{Forman}}},\
  }\href {\doibase 10.1088/0004-637X/710/2/1776} {\bibfield  {journal}
  {\bibinfo  {journal} {\apj}\ }\textbf {\bibinfo {volume} {710}},\ \bibinfo
  {pages} {1776} (\bibinfo {year} {2010})},\ \Eprint
  {http://arxiv.org/abs/1001.2441} {arXiv:1001.2441 [astro-ph.CO]} \BibitemShut
  {NoStop}%
\bibitem [{\citenamefont {{Randall}}\ \emph {et~al.}(2010)\citenamefont
  {{Randall}}, \citenamefont {{Clarke}}, \citenamefont {{Nulsen}},
  \citenamefont {{Owers}}, \citenamefont {{Sarazin}}, \citenamefont
  {{Forman}},\ and\ \citenamefont {{Murray}}}]{randall10}%
  \BibitemOpen
  \bibfield  {author} {\bibinfo {author} {\bibfnamefont {S.~W.}\ \bibnamefont
  {{Randall}}}, \bibinfo {author} {\bibfnamefont {T.~E.}\ \bibnamefont
  {{Clarke}}}, \bibinfo {author} {\bibfnamefont {P.~E.~J.}\ \bibnamefont
  {{Nulsen}}}, \bibinfo {author} {\bibfnamefont {M.~S.}\ \bibnamefont
  {{Owers}}}, \bibinfo {author} {\bibfnamefont {C.~L.}\ \bibnamefont
  {{Sarazin}}}, \bibinfo {author} {\bibfnamefont {W.~R.}\ \bibnamefont
  {{Forman}}}, \ and\ \bibinfo {author} {\bibfnamefont {S.~S.}\ \bibnamefont
  {{Murray}}},\ }\href {\doibase 10.1088/0004-637X/722/1/825} {\bibfield
  {journal} {\bibinfo  {journal} {\apj}\ }\textbf {\bibinfo {volume} {722}},\
  \bibinfo {pages} {825} (\bibinfo {year} {2010})},\ \Eprint
  {http://arxiv.org/abs/1008.2921} {arXiv:1008.2921 [astro-ph.CO]} \BibitemShut
  {NoStop}%
\bibitem [{\citenamefont {{Johnson}}\ \emph {et~al.}(2012)\citenamefont
  {{Johnson}}, \citenamefont {{Zuhone}}, \citenamefont {{Jones}}, \citenamefont
  {{Forman}},\ and\ \citenamefont {{Markevitch}}}]{johnson12}%
  \BibitemOpen
  \bibfield  {author} {\bibinfo {author} {\bibfnamefont {R.~E.}\ \bibnamefont
  {{Johnson}}}, \bibinfo {author} {\bibfnamefont {J.}~\bibnamefont {{Zuhone}}},
  \bibinfo {author} {\bibfnamefont {C.}~\bibnamefont {{Jones}}}, \bibinfo
  {author} {\bibfnamefont {W.~R.}\ \bibnamefont {{Forman}}}, \ and\ \bibinfo
  {author} {\bibfnamefont {M.}~\bibnamefont {{Markevitch}}},\ }\href {\doibase
  10.1088/0004-637X/751/2/95} {\bibfield  {journal} {\bibinfo  {journal}
  {\apj}\ }\textbf {\bibinfo {volume} {751}},\ \bibinfo {eid} {95} (\bibinfo
  {year} {2012})},\ \Eprint {http://arxiv.org/abs/1106.3489} {arXiv:1106.3489
  [astro-ph.CO]} \BibitemShut {NoStop}%
\bibitem [{\citenamefont {{Reiprich}}\ and\ \citenamefont
  {{B{\"o}hringer}}(2002)}]{reiprich02}%
  \BibitemOpen
  \bibfield  {author} {\bibinfo {author} {\bibfnamefont {T.~H.}\ \bibnamefont
  {{Reiprich}}}\ and\ \bibinfo {author} {\bibfnamefont {H.}~\bibnamefont
  {{B{\"o}hringer}}},\ }\href {\doibase 10.1086/338753} {\bibfield  {journal}
  {\bibinfo  {journal} {\apj}\ }\textbf {\bibinfo {volume} {567}},\ \bibinfo
  {pages} {716} (\bibinfo {year} {2002})},\ \Eprint
  {http://arxiv.org/abs/arXiv:astro-ph/0111285} {arXiv:astro-ph/0111285}
  \BibitemShut {NoStop}%
\bibitem [{\citenamefont {{Babyk}}\ \emph {et~al.}(2012)\citenamefont
  {{Babyk}}, \citenamefont {{Vavilova}},\ and\ \citenamefont {{Del
  Popolo}}}]{babyk12}%
  \BibitemOpen
  \bibfield  {author} {\bibinfo {author} {\bibfnamefont {I.}~\bibnamefont
  {{Babyk}}}, \bibinfo {author} {\bibfnamefont {I.}~\bibnamefont {{Vavilova}}},
  \ and\ \bibinfo {author} {\bibfnamefont {A.}~\bibnamefont {{Del Popolo}}},\
  }\href@noop {} {\bibfield  {journal} {\bibinfo  {journal} {ArXiv e-prints}\ }
  (\bibinfo {year} {2012})},\ \Eprint {http://arxiv.org/abs/1208.2424}
  {arXiv:1208.2424 [astro-ph.CO]} \BibitemShut {NoStop}%
\bibitem [{\citenamefont {{Pimbblet}}\ \emph {et~al.}(2006)\citenamefont
  {{Pimbblet}}, \citenamefont {{Smail}}, \citenamefont {{Edge}}, \citenamefont
  {{O'Hely}}, \citenamefont {{Couch}},\ and\ \citenamefont
  {{Zabludoff}}}]{pimbblet06}%
  \BibitemOpen
  \bibfield  {author} {\bibinfo {author} {\bibfnamefont {K.~A.}\ \bibnamefont
  {{Pimbblet}}}, \bibinfo {author} {\bibfnamefont {I.}~\bibnamefont {{Smail}}},
  \bibinfo {author} {\bibfnamefont {A.~C.}\ \bibnamefont {{Edge}}}, \bibinfo
  {author} {\bibfnamefont {E.}~\bibnamefont {{O'Hely}}}, \bibinfo {author}
  {\bibfnamefont {W.~J.}\ \bibnamefont {{Couch}}}, \ and\ \bibinfo {author}
  {\bibfnamefont {A.~I.}\ \bibnamefont {{Zabludoff}}},\ }\href {\doibase
  10.1111/j.1365-2966.2005.09892.x} {\bibfield  {journal} {\bibinfo  {journal}
  {\mnras}\ }\textbf {\bibinfo {volume} {366}},\ \bibinfo {pages} {645}
  (\bibinfo {year} {2006})},\ \Eprint {http://arxiv.org/abs/astro-ph/0511667}
  {astro-ph/0511667} \BibitemShut {NoStop}%
\bibitem [{\citenamefont {{Markevitch}}\ \emph {et~al.}(2000)\citenamefont
  {{Markevitch}}, \citenamefont {{Ponman}}, \citenamefont {{Nulsen}},
  \citenamefont {{Bautz}}, \citenamefont {{Burke}}, \citenamefont {{David}},
  \citenamefont {{Davis}}, \citenamefont {{Donnelly}}, \citenamefont
  {{Forman}}, \citenamefont {{Jones}}, \citenamefont {{Kaastra}}, \citenamefont
  {{Kellogg}}, \citenamefont {{Kim}}, \citenamefont {{Kolodziejczak}},
  \citenamefont {{Mazzotta}}, \citenamefont {{Pagliaro}}, \citenamefont
  {{Patel}}, \citenamefont {{Van Speybroeck}}, \citenamefont {{Vikhlinin}},
  \citenamefont {{Vrtilek}}, \citenamefont {{Wise}},\ and\ \citenamefont
  {{Zhao}}}]{Markevitch00}%
  \BibitemOpen
  \bibfield  {author} {\bibinfo {author} {\bibfnamefont {M.}~\bibnamefont
  {{Markevitch}}}, \bibinfo {author} {\bibfnamefont {T.~J.}\ \bibnamefont
  {{Ponman}}}, \bibinfo {author} {\bibfnamefont {P.~E.~J.}\ \bibnamefont
  {{Nulsen}}}, \bibinfo {author} {\bibfnamefont {M.~W.}\ \bibnamefont
  {{Bautz}}}, \bibinfo {author} {\bibfnamefont {D.~J.}\ \bibnamefont
  {{Burke}}}, \bibinfo {author} {\bibfnamefont {L.~P.}\ \bibnamefont
  {{David}}}, \bibinfo {author} {\bibfnamefont {D.}~\bibnamefont {{Davis}}},
  \bibinfo {author} {\bibfnamefont {R.~H.}\ \bibnamefont {{Donnelly}}},
  \bibinfo {author} {\bibfnamefont {W.~R.}\ \bibnamefont {{Forman}}}, \bibinfo
  {author} {\bibfnamefont {C.}~\bibnamefont {{Jones}}}, \bibinfo {author}
  {\bibfnamefont {J.}~\bibnamefont {{Kaastra}}}, \bibinfo {author}
  {\bibfnamefont {E.}~\bibnamefont {{Kellogg}}}, \bibinfo {author}
  {\bibfnamefont {D.-W.}\ \bibnamefont {{Kim}}}, \bibinfo {author}
  {\bibfnamefont {J.}~\bibnamefont {{Kolodziejczak}}}, \bibinfo {author}
  {\bibfnamefont {P.}~\bibnamefont {{Mazzotta}}}, \bibinfo {author}
  {\bibfnamefont {A.}~\bibnamefont {{Pagliaro}}}, \bibinfo {author}
  {\bibfnamefont {S.}~\bibnamefont {{Patel}}}, \bibinfo {author} {\bibfnamefont
  {L.}~\bibnamefont {{Van Speybroeck}}}, \bibinfo {author} {\bibfnamefont
  {A.}~\bibnamefont {{Vikhlinin}}}, \bibinfo {author} {\bibfnamefont
  {J.}~\bibnamefont {{Vrtilek}}}, \bibinfo {author} {\bibfnamefont
  {M.}~\bibnamefont {{Wise}}}, \ and\ \bibinfo {author} {\bibfnamefont
  {P.}~\bibnamefont {{Zhao}}},\ }\href {\doibase 10.1086/309470} {\bibfield
  {journal} {\bibinfo  {journal} {\apj}\ }\textbf {\bibinfo {volume} {541}},\
  \bibinfo {pages} {542} (\bibinfo {year} {2000})},\ \Eprint
  {http://arxiv.org/abs/arXiv:astro-ph/0001269} {arXiv:astro-ph/0001269}
  \BibitemShut {NoStop}%
\bibitem [{\citenamefont {{Russell}}\ \emph {et~al.}(2008)\citenamefont
  {{Russell}}, \citenamefont {{Sanders}},\ and\ \citenamefont
  {{Fabian}}}]{russell08}%
  \BibitemOpen
  \bibfield  {author} {\bibinfo {author} {\bibfnamefont {H.~R.}\ \bibnamefont
  {{Russell}}}, \bibinfo {author} {\bibfnamefont {J.~S.}\ \bibnamefont
  {{Sanders}}}, \ and\ \bibinfo {author} {\bibfnamefont {A.~C.}\ \bibnamefont
  {{Fabian}}},\ }\href {\doibase 10.1111/j.1365-2966.2008.13823.x} {\bibfield
  {journal} {\bibinfo  {journal} {\mnras}\ }\textbf {\bibinfo {volume} {390}},\
  \bibinfo {pages} {1207} (\bibinfo {year} {2008})},\ \Eprint
  {http://arxiv.org/abs/0808.2371} {arXiv:0808.2371} \BibitemShut {NoStop}%
\bibitem [{\citenamefont {{Allen}}\ \emph {et~al.}(2001)\citenamefont
  {{Allen}}, \citenamefont {{Ettori}},\ and\ \citenamefont
  {{Fabian}}}]{allen01}%
  \BibitemOpen
  \bibfield  {author} {\bibinfo {author} {\bibfnamefont {S.~W.}\ \bibnamefont
  {{Allen}}}, \bibinfo {author} {\bibfnamefont {S.}~\bibnamefont {{Ettori}}}, \
  and\ \bibinfo {author} {\bibfnamefont {A.~C.}\ \bibnamefont {{Fabian}}},\
  }\href {\doibase 10.1046/j.1365-8711.2001.04318.x} {\bibfield  {journal}
  {\bibinfo  {journal} {\mnras}\ }\textbf {\bibinfo {volume} {324}},\ \bibinfo
  {pages} {877} (\bibinfo {year} {2001})},\ \Eprint
  {http://arxiv.org/abs/arXiv:astro-ph/0008517} {arXiv:astro-ph/0008517}
  \BibitemShut {NoStop}%
\bibitem [{\citenamefont {{Lal}}\ \emph {et~al.}(2013)\citenamefont {{Lal}},
  \citenamefont {{Kraft}}, \citenamefont {{Randall}}, \citenamefont {{Forman}},
  \citenamefont {{Nulsen}}, \citenamefont {{Roediger}}, \citenamefont
  {{ZuHone}}, \citenamefont {{Hardcastle}}, \citenamefont {{Jones}},\ and\
  \citenamefont {{Croston}}}]{lal13}%
  \BibitemOpen
  \bibfield  {author} {\bibinfo {author} {\bibfnamefont {D.~V.}\ \bibnamefont
  {{Lal}}}, \bibinfo {author} {\bibfnamefont {R.~P.}\ \bibnamefont {{Kraft}}},
  \bibinfo {author} {\bibfnamefont {S.~W.}\ \bibnamefont {{Randall}}}, \bibinfo
  {author} {\bibfnamefont {W.~R.}\ \bibnamefont {{Forman}}}, \bibinfo {author}
  {\bibfnamefont {P.~E.~J.}\ \bibnamefont {{Nulsen}}}, \bibinfo {author}
  {\bibfnamefont {E.}~\bibnamefont {{Roediger}}}, \bibinfo {author}
  {\bibfnamefont {J.~A.}\ \bibnamefont {{ZuHone}}}, \bibinfo {author}
  {\bibfnamefont {M.~J.}\ \bibnamefont {{Hardcastle}}}, \bibinfo {author}
  {\bibfnamefont {C.}~\bibnamefont {{Jones}}}, \ and\ \bibinfo {author}
  {\bibfnamefont {J.~H.}\ \bibnamefont {{Croston}}},\ }\href {\doibase
  10.1088/0004-637X/764/1/83} {\bibfield  {journal} {\bibinfo  {journal}
  {\apj}\ }\textbf {\bibinfo {volume} {764}},\ \bibinfo {eid} {83} (\bibinfo
  {year} {2013})},\ \Eprint {http://arxiv.org/abs/1210.7563} {arXiv:1210.7563
  [astro-ph.CO]} \BibitemShut {NoStop}%
\bibitem [{\citenamefont {{Gastaldello}}\ \emph {et~al.}(2013)\citenamefont
  {{Gastaldello}}, \citenamefont {{Di Gesu}}, \citenamefont {{Ghizzardi}},
  \citenamefont {{Giacintucci}}, \citenamefont {{Girardi}}, \citenamefont
  {{Roediger}}, \citenamefont {{Rossetti}}, \citenamefont {{Brighenti}},
  \citenamefont {{Buote}}, \citenamefont {{Eckert}}, \citenamefont {{Ettori}},
  \citenamefont {{Humphrey}},\ and\ \citenamefont {{Mathews}}}]{gastaldello13}%
  \BibitemOpen
  \bibfield  {author} {\bibinfo {author} {\bibfnamefont {F.}~\bibnamefont
  {{Gastaldello}}}, \bibinfo {author} {\bibfnamefont {L.}~\bibnamefont {{Di
  Gesu}}}, \bibinfo {author} {\bibfnamefont {S.}~\bibnamefont {{Ghizzardi}}},
  \bibinfo {author} {\bibfnamefont {S.}~\bibnamefont {{Giacintucci}}}, \bibinfo
  {author} {\bibfnamefont {M.}~\bibnamefont {{Girardi}}}, \bibinfo {author}
  {\bibfnamefont {E.}~\bibnamefont {{Roediger}}}, \bibinfo {author}
  {\bibfnamefont {M.}~\bibnamefont {{Rossetti}}}, \bibinfo {author}
  {\bibfnamefont {F.}~\bibnamefont {{Brighenti}}}, \bibinfo {author}
  {\bibfnamefont {D.~A.}\ \bibnamefont {{Buote}}}, \bibinfo {author}
  {\bibfnamefont {D.}~\bibnamefont {{Eckert}}}, \bibinfo {author}
  {\bibfnamefont {S.}~\bibnamefont {{Ettori}}}, \bibinfo {author}
  {\bibfnamefont {P.~J.}\ \bibnamefont {{Humphrey}}}, \ and\ \bibinfo {author}
  {\bibfnamefont {W.~G.}\ \bibnamefont {{Mathews}}},\ }\href@noop {} {\bibfield
   {journal} {\bibinfo  {journal} {ArXiv e-prints}\ } (\bibinfo {year}
  {2013})},\ \Eprint {http://arxiv.org/abs/1304.5478} {arXiv:1304.5478
  [astro-ph.CO]} \BibitemShut {NoStop}%
\bibitem [{\citenamefont {{Graham}}\ \emph {et~al.}(2008)\citenamefont
  {{Graham}}, \citenamefont {{Fabian}},\ and\ \citenamefont
  {{Sanders}}}]{GrahamEtAl08}%
  \BibitemOpen
  \bibfield  {author} {\bibinfo {author} {\bibfnamefont {J.}~\bibnamefont
  {{Graham}}}, \bibinfo {author} {\bibfnamefont {A.~C.}\ \bibnamefont
  {{Fabian}}}, \ and\ \bibinfo {author} {\bibfnamefont {J.~S.}\ \bibnamefont
  {{Sanders}}},\ }\href {\doibase 10.1111/j.1365-2966.2008.13027.x} {\bibfield
  {journal} {\bibinfo  {journal} {\mnras}\ }\textbf {\bibinfo {volume} {386}},\
  \bibinfo {pages} {278} (\bibinfo {year} {2008})},\ \Eprint
  {http://arxiv.org/abs/0801.4253} {arXiv:0801.4253} \BibitemShut {NoStop}%
\bibitem [{\citenamefont {{Blanton}}\ \emph {et~al.}(2011)\citenamefont
  {{Blanton}}, \citenamefont {{Randall}}, \citenamefont {{Clarke}},
  \citenamefont {{Sarazin}}, \citenamefont {{McNamara}}, \citenamefont
  {{Douglass}},\ and\ \citenamefont {{McDonald}}}]{blanton11}%
  \BibitemOpen
  \bibfield  {author} {\bibinfo {author} {\bibfnamefont {E.~L.}\ \bibnamefont
  {{Blanton}}}, \bibinfo {author} {\bibfnamefont {S.~W.}\ \bibnamefont
  {{Randall}}}, \bibinfo {author} {\bibfnamefont {T.~E.}\ \bibnamefont
  {{Clarke}}}, \bibinfo {author} {\bibfnamefont {C.~L.}\ \bibnamefont
  {{Sarazin}}}, \bibinfo {author} {\bibfnamefont {B.~R.}\ \bibnamefont
  {{McNamara}}}, \bibinfo {author} {\bibfnamefont {E.~M.}\ \bibnamefont
  {{Douglass}}}, \ and\ \bibinfo {author} {\bibfnamefont {M.}~\bibnamefont
  {{McDonald}}},\ }\href {\doibase 10.1088/0004-637X/737/2/99} {\bibfield
  {journal} {\bibinfo  {journal} {\apj}\ }\textbf {\bibinfo {volume} {737}},\
  \bibinfo {eid} {99} (\bibinfo {year} {2011})},\ \Eprint
  {http://arxiv.org/abs/1105.4572} {arXiv:1105.4572 [astro-ph.CO]} \BibitemShut
  {NoStop}%
\bibitem [{\citenamefont {{Mazzotta}}\ \emph {et~al.}(2004)\citenamefont
  {{Mazzotta}}, \citenamefont {{Rasia}}, \citenamefont {{Moscardini}},\ and\
  \citenamefont {{Tormen}}}]{MazzottaDeprojection}%
  \BibitemOpen
  \bibfield  {author} {\bibinfo {author} {\bibfnamefont {P.}~\bibnamefont
  {{Mazzotta}}}, \bibinfo {author} {\bibfnamefont {E.}~\bibnamefont {{Rasia}}},
  \bibinfo {author} {\bibfnamefont {L.}~\bibnamefont {{Moscardini}}}, \ and\
  \bibinfo {author} {\bibfnamefont {G.}~\bibnamefont {{Tormen}}},\ }\href
  {\doibase 10.1111/j.1365-2966.2004.08167.x} {\bibfield  {journal} {\bibinfo
  {journal} {\mnras}\ }\textbf {\bibinfo {volume} {354}},\ \bibinfo {pages}
  {10} (\bibinfo {year} {2004})},\ \Eprint
  {http://arxiv.org/abs/astro-ph/0409618} {astro-ph/0409618} \BibitemShut
  {NoStop}%
\bibitem [{\citenamefont {{Russell}}\ \emph {et~al.}(2010)\citenamefont
  {{Russell}}, \citenamefont {{Sanders}}, \citenamefont {{Fabian}},
  \citenamefont {{Baum}}, \citenamefont {{Donahue}}, \citenamefont {{Edge}},
  \citenamefont {{McNamara}},\ and\ \citenamefont {{O'Dea}}}]{russell10}%
  \BibitemOpen
  \bibfield  {author} {\bibinfo {author} {\bibfnamefont {H.~R.}\ \bibnamefont
  {{Russell}}}, \bibinfo {author} {\bibfnamefont {J.~S.}\ \bibnamefont
  {{Sanders}}}, \bibinfo {author} {\bibfnamefont {A.~C.}\ \bibnamefont
  {{Fabian}}}, \bibinfo {author} {\bibfnamefont {S.~A.}\ \bibnamefont
  {{Baum}}}, \bibinfo {author} {\bibfnamefont {M.}~\bibnamefont {{Donahue}}},
  \bibinfo {author} {\bibfnamefont {A.~C.}\ \bibnamefont {{Edge}}}, \bibinfo
  {author} {\bibfnamefont {B.~R.}\ \bibnamefont {{McNamara}}}, \ and\ \bibinfo
  {author} {\bibfnamefont {C.~P.}\ \bibnamefont {{O'Dea}}},\ }\href {\doibase
  10.1111/j.1365-2966.2010.16822.x} {\bibfield  {journal} {\bibinfo  {journal}
  {\mnras}\ }\textbf {\bibinfo {volume} {406}},\ \bibinfo {pages} {1721}
  (\bibinfo {year} {2010})},\ \Eprint {http://arxiv.org/abs/1004.1559}
  {arXiv:1004.1559 [astro-ph.CO]} \BibitemShut {NoStop}%
\bibitem [{\citenamefont {{Forman}}\ \emph {et~al.}(2007)\citenamefont
  {{Forman}}, \citenamefont {{Jones}}, \citenamefont {{Churazov}},
  \citenamefont {{Markevitch}}, \citenamefont {{Nulsen}}, \citenamefont
  {{Vikhlinin}}, \citenamefont {{Begelman}}, \citenamefont {{B{\"o}hringer}},
  \citenamefont {{Eilek}}, \citenamefont {{Heinz}}, \citenamefont {{Kraft}},
  \citenamefont {{Owen}},\ and\ \citenamefont {{Pahre}}}]{forman07}%
  \BibitemOpen
  \bibfield  {author} {\bibinfo {author} {\bibfnamefont {W.}~\bibnamefont
  {{Forman}}}, \bibinfo {author} {\bibfnamefont {C.}~\bibnamefont {{Jones}}},
  \bibinfo {author} {\bibfnamefont {E.}~\bibnamefont {{Churazov}}}, \bibinfo
  {author} {\bibfnamefont {M.}~\bibnamefont {{Markevitch}}}, \bibinfo {author}
  {\bibfnamefont {P.}~\bibnamefont {{Nulsen}}}, \bibinfo {author}
  {\bibfnamefont {A.}~\bibnamefont {{Vikhlinin}}}, \bibinfo {author}
  {\bibfnamefont {M.}~\bibnamefont {{Begelman}}}, \bibinfo {author}
  {\bibfnamefont {H.}~\bibnamefont {{B{\"o}hringer}}}, \bibinfo {author}
  {\bibfnamefont {J.}~\bibnamefont {{Eilek}}}, \bibinfo {author} {\bibfnamefont
  {S.}~\bibnamefont {{Heinz}}}, \bibinfo {author} {\bibfnamefont
  {R.}~\bibnamefont {{Kraft}}}, \bibinfo {author} {\bibfnamefont
  {F.}~\bibnamefont {{Owen}}}, \ and\ \bibinfo {author} {\bibfnamefont
  {M.}~\bibnamefont {{Pahre}}},\ }\href {\doibase 10.1086/519480} {\bibfield
  {journal} {\bibinfo  {journal} {\apj}\ }\textbf {\bibinfo {volume} {665}},\
  \bibinfo {pages} {1057} (\bibinfo {year} {2007})},\ \Eprint
  {http://arxiv.org/abs/arXiv:astro-ph/0604583} {arXiv:astro-ph/0604583}
  \BibitemShut {NoStop}%
\bibitem [{\citenamefont {{Markevitch}}(2006)}]{markevitch05}%
  \BibitemOpen
  \bibfield  {author} {\bibinfo {author} {\bibfnamefont {M.}~\bibnamefont
  {{Markevitch}}},\ }in\ \href@noop {} {\emph {\bibinfo {booktitle} {The X-ray
  Universe 2005}}},\ \bibinfo {series} {ESA Special Publication}, Vol.\
  \bibinfo {volume} {604},\ \bibinfo {editor} {edited by\ \bibinfo {editor}
  {\bibfnamefont {A.}~\bibnamefont {{Wilson}}}}\ (\bibinfo {year} {2006})\ p.\
  \bibinfo {pages} {723},\ \Eprint
  {http://arxiv.org/abs/arXiv:astro-ph/0511345} {arXiv:astro-ph/0511345}
  \BibitemShut {NoStop}%
\bibitem [{\citenamefont {{Russell}}\ \emph {et~al.}(2012)\citenamefont
  {{Russell}}, \citenamefont {{McNamara}}, \citenamefont {{Sanders}},
  \citenamefont {{Fabian}}, \citenamefont {{Nulsen}}, \citenamefont
  {{Canning}}, \citenamefont {{Baum}}, \citenamefont {{Donahue}}, \citenamefont
  {{Edge}}, \citenamefont {{King}},\ and\ \citenamefont {{O'Dea}}}]{russell12}%
  \BibitemOpen
  \bibfield  {author} {\bibinfo {author} {\bibfnamefont {H.~R.}\ \bibnamefont
  {{Russell}}}, \bibinfo {author} {\bibfnamefont {B.~R.}\ \bibnamefont
  {{McNamara}}}, \bibinfo {author} {\bibfnamefont {J.~S.}\ \bibnamefont
  {{Sanders}}}, \bibinfo {author} {\bibfnamefont {A.~C.}\ \bibnamefont
  {{Fabian}}}, \bibinfo {author} {\bibfnamefont {P.~E.~J.}\ \bibnamefont
  {{Nulsen}}}, \bibinfo {author} {\bibfnamefont {R.~E.~A.}\ \bibnamefont
  {{Canning}}}, \bibinfo {author} {\bibfnamefont {S.~A.}\ \bibnamefont
  {{Baum}}}, \bibinfo {author} {\bibfnamefont {M.}~\bibnamefont {{Donahue}}},
  \bibinfo {author} {\bibfnamefont {A.~C.}\ \bibnamefont {{Edge}}}, \bibinfo
  {author} {\bibfnamefont {L.~J.}\ \bibnamefont {{King}}}, \ and\ \bibinfo
  {author} {\bibfnamefont {C.~P.}\ \bibnamefont {{O'Dea}}},\ }\href {\doibase
  10.1111/j.1365-2966.2012.20808.x} {\bibfield  {journal} {\bibinfo  {journal}
  {\mnras}\ }\textbf {\bibinfo {volume} {423}},\ \bibinfo {pages} {236}
  (\bibinfo {year} {2012})},\ \Eprint {http://arxiv.org/abs/1202.5320}
  {arXiv:1202.5320 [astro-ph.CO]} \BibitemShut {NoStop}%
\bibitem [{\citenamefont {{Churazov}}\ \emph {et~al.}(2008)\citenamefont
  {{Churazov}}, \citenamefont {{Forman}}, \citenamefont {{Vikhlinin}},
  \citenamefont {{Tremaine}}, \citenamefont {{Gerhard}},\ and\ \citenamefont
  {{Jones}}}]{churazov08}%
  \BibitemOpen
  \bibfield  {author} {\bibinfo {author} {\bibfnamefont {E.}~\bibnamefont
  {{Churazov}}}, \bibinfo {author} {\bibfnamefont {W.}~\bibnamefont
  {{Forman}}}, \bibinfo {author} {\bibfnamefont {A.}~\bibnamefont
  {{Vikhlinin}}}, \bibinfo {author} {\bibfnamefont {S.}~\bibnamefont
  {{Tremaine}}}, \bibinfo {author} {\bibfnamefont {O.}~\bibnamefont
  {{Gerhard}}}, \ and\ \bibinfo {author} {\bibfnamefont {C.}~\bibnamefont
  {{Jones}}},\ }\href {\doibase 10.1111/j.1365-2966.2008.13507.x} {\bibfield
  {journal} {\bibinfo  {journal} {\mnras}\ }\textbf {\bibinfo {volume} {388}},\
  \bibinfo {pages} {1062} (\bibinfo {year} {2008})},\ \Eprint
  {http://arxiv.org/abs/0711.4686} {arXiv:0711.4686} \BibitemShut {NoStop}%
\bibitem [{\citenamefont {{Chiu}}\ \emph {et~al.}(2012)\citenamefont {{Chiu}},
  \citenamefont {{Molnar}},\ and\ \citenamefont {{Chen}}}]{chiu12}%
  \BibitemOpen
  \bibfield  {author} {\bibinfo {author} {\bibfnamefont {I.}~\bibnamefont
  {{Chiu}}}, \bibinfo {author} {\bibfnamefont {S.~M.}\ \bibnamefont
  {{Molnar}}}, \ and\ \bibinfo {author} {\bibfnamefont {P.}~\bibnamefont
  {{Chen}}},\ }\href@noop {} {\bibfield  {journal} {\bibinfo  {journal} {ArXiv
  e-prints}\ } (\bibinfo {year} {2012})},\ \Eprint
  {http://arxiv.org/abs/1206.0532} {arXiv:1206.0532 [astro-ph.CO]} \BibitemShut
  {NoStop}%
\bibitem [{\citenamefont {{Parrish}}\ and\ \citenamefont
  {{Quataert}}(2008)}]{parrish08}%
  \BibitemOpen
  \bibfield  {author} {\bibinfo {author} {\bibfnamefont {I.~J.}\ \bibnamefont
  {{Parrish}}}\ and\ \bibinfo {author} {\bibfnamefont {E.}~\bibnamefont
  {{Quataert}}},\ }\href {\doibase 10.1086/587937} {\bibfield  {journal}
  {\bibinfo  {journal} {\apjl}\ }\textbf {\bibinfo {volume} {677}},\ \bibinfo
  {pages} {L9} (\bibinfo {year} {2008})},\ \Eprint
  {http://arxiv.org/abs/0712.3048} {arXiv:0712.3048} \BibitemShut {NoStop}%
\bibitem [{\citenamefont {{Vikhlinin}}\ and\ \citenamefont
  {{Markevitch}}(2002)}]{vikhlinin02}%
  \BibitemOpen
  \bibfield  {author} {\bibinfo {author} {\bibfnamefont {A.~A.}\ \bibnamefont
  {{Vikhlinin}}}\ and\ \bibinfo {author} {\bibfnamefont {M.~L.}\ \bibnamefont
  {{Markevitch}}},\ }\href {\doibase 10.1134/1.1499173} {\bibfield  {journal}
  {\bibinfo  {journal} {Astronomy Letters}\ }\textbf {\bibinfo {volume} {28}},\
  \bibinfo {pages} {495} (\bibinfo {year} {2002})},\ \Eprint
  {http://arxiv.org/abs/arXiv:astro-ph/0209551} {arXiv:astro-ph/0209551}
  \BibitemShut {NoStop}%
\bibitem [{\citenamefont {{Keshet}}(2010)}]{keshet10relics}%
  \BibitemOpen
  \bibfield  {author} {\bibinfo {author} {\bibfnamefont {U.}~\bibnamefont
  {{Keshet}}},\ }\href@noop {} {\bibfield  {journal} {\bibinfo  {journal}
  {ArXiv e-prints}\ } (\bibinfo {year} {2010})},\ \Eprint
  {http://arxiv.org/abs/1011.0729} {arXiv:1011.0729 [astro-ph.HE]} \BibitemShut
  {NoStop}%
\bibitem [{\citenamefont {{Sanders}}\ \emph {et~al.}(2010)\citenamefont
  {{Sanders}}, \citenamefont {{Fabian}}, \citenamefont {{Smith}},\ and\
  \citenamefont {{Peterson}}}]{sanders10turbulence}%
  \BibitemOpen
  \bibfield  {author} {\bibinfo {author} {\bibfnamefont {J.~S.}\ \bibnamefont
  {{Sanders}}}, \bibinfo {author} {\bibfnamefont {A.~C.}\ \bibnamefont
  {{Fabian}}}, \bibinfo {author} {\bibfnamefont {R.~K.}\ \bibnamefont
  {{Smith}}}, \ and\ \bibinfo {author} {\bibfnamefont {J.~R.}\ \bibnamefont
  {{Peterson}}},\ }\href {\doibase 10.1111/j.1745-3933.2009.00789.x} {\bibfield
   {journal} {\bibinfo  {journal} {\mnras}\ }\textbf {\bibinfo {volume}
  {402}},\ \bibinfo {pages} {L11} (\bibinfo {year} {2010})},\ \Eprint
  {http://arxiv.org/abs/0911.0763} {arXiv:0911.0763 [astro-ph.CO]} \BibitemShut
  {NoStop}%
\bibitem [{\citenamefont {{Sanders}}\ \emph {et~al.}(2011)\citenamefont
  {{Sanders}}, \citenamefont {{Fabian}},\ and\ \citenamefont
  {{Smith}}}]{sanders11turbulence}%
  \BibitemOpen
  \bibfield  {author} {\bibinfo {author} {\bibfnamefont {J.~S.}\ \bibnamefont
  {{Sanders}}}, \bibinfo {author} {\bibfnamefont {A.~C.}\ \bibnamefont
  {{Fabian}}}, \ and\ \bibinfo {author} {\bibfnamefont {R.~K.}\ \bibnamefont
  {{Smith}}},\ }\href {\doibase 10.1111/j.1365-2966.2010.17561.x} {\bibfield
  {journal} {\bibinfo  {journal} {\mnras}\ }\textbf {\bibinfo {volume} {410}},\
  \bibinfo {pages} {1797} (\bibinfo {year} {2011})},\ \Eprint
  {http://arxiv.org/abs/1008.3500} {arXiv:1008.3500 [astro-ph.CO]} \BibitemShut
  {NoStop}%
\bibitem [{\citenamefont {{Bulbul}}\ \emph {et~al.}(2012)\citenamefont
  {{Bulbul}}, \citenamefont {{Smith}}, \citenamefont {{Foster}}, \citenamefont
  {{Cottam}}, \citenamefont {{Loewenstein}}, \citenamefont {{Mushotzky}},\ and\
  \citenamefont {{Shafer}}}]{bulbul12}%
  \BibitemOpen
  \bibfield  {author} {\bibinfo {author} {\bibfnamefont {G.~E.}\ \bibnamefont
  {{Bulbul}}}, \bibinfo {author} {\bibfnamefont {R.~K.}\ \bibnamefont
  {{Smith}}}, \bibinfo {author} {\bibfnamefont {A.}~\bibnamefont {{Foster}}},
  \bibinfo {author} {\bibfnamefont {J.}~\bibnamefont {{Cottam}}}, \bibinfo
  {author} {\bibfnamefont {M.}~\bibnamefont {{Loewenstein}}}, \bibinfo {author}
  {\bibfnamefont {R.}~\bibnamefont {{Mushotzky}}}, \ and\ \bibinfo {author}
  {\bibfnamefont {R.}~\bibnamefont {{Shafer}}},\ }\href {\doibase
  10.1088/0004-637X/747/1/32} {\bibfield  {journal} {\bibinfo  {journal}
  {\apj}\ }\textbf {\bibinfo {volume} {747}},\ \bibinfo {eid} {32} (\bibinfo
  {year} {2012})},\ \Eprint {http://arxiv.org/abs/1110.4422} {arXiv:1110.4422
  [astro-ph.CO]} \BibitemShut {NoStop}%
\bibitem [{\citenamefont {{Molendi}}\ and\ \citenamefont
  {{Pizzolato}}(2001)}]{molendi01}%
  \BibitemOpen
  \bibfield  {author} {\bibinfo {author} {\bibfnamefont {S.}~\bibnamefont
  {{Molendi}}}\ and\ \bibinfo {author} {\bibfnamefont {F.}~\bibnamefont
  {{Pizzolato}}},\ }\href {\doibase 10.1086/322387} {\bibfield  {journal}
  {\bibinfo  {journal} {\apj}\ }\textbf {\bibinfo {volume} {560}},\ \bibinfo
  {pages} {194} (\bibinfo {year} {2001})},\ \Eprint
  {http://arxiv.org/abs/arXiv:astro-ph/0106552} {arXiv:astro-ph/0106552}
  \BibitemShut {NoStop}%
\bibitem [{\citenamefont {{Ghizzardi}}\ \emph {et~al.}(2012)\citenamefont
  {{Ghizzardi}}, \citenamefont {{De Grandi}},\ and\ \citenamefont
  {{Molendi}}}]{ghizzardi12}%
  \BibitemOpen
  \bibfield  {author} {\bibinfo {author} {\bibfnamefont {S.}~\bibnamefont
  {{Ghizzardi}}}, \bibinfo {author} {\bibfnamefont {S.}~\bibnamefont {{De
  Grandi}}}, \ and\ \bibinfo {author} {\bibfnamefont {S.}~\bibnamefont
  {{Molendi}}},\ }in\ \href@noop {} {\emph {\bibinfo {booktitle} {Galaxy
  Clusters as Giant Cosmic Laboratories, Proceedings of a workshop held 21-23
  May 2012 in Madrid, Spain. Organized by the XMM-Newton Science Operations
  Centre of the European Space Agency (ESA)., p.16}}},\ \bibinfo {editor}
  {edited by\ \bibinfo {editor} {\bibfnamefont {J.-U.}\ \bibnamefont
  {{Ness}}}}\ (\bibinfo {year} {2012})\ p.~\bibinfo {pages} {16}\BibitemShut
  {NoStop}%
\bibitem [{\citenamefont {{Kushnir}}\ \emph {et~al.}(2009)\citenamefont
  {{Kushnir}}, \citenamefont {{Katz}},\ and\ \citenamefont
  {{Waxman}}}]{Kushnir2009}%
  \BibitemOpen
  \bibfield  {author} {\bibinfo {author} {\bibfnamefont {D.}~\bibnamefont
  {{Kushnir}}}, \bibinfo {author} {\bibfnamefont {B.}~\bibnamefont {{Katz}}}, \
  and\ \bibinfo {author} {\bibfnamefont {E.}~\bibnamefont {{Waxman}}},\ }\href
  {\doibase 10.1088/1475-7516/2009/09/024} {\bibfield  {journal} {\bibinfo
  {journal} {\jcap}\ }\textbf {\bibinfo {volume} {9}},\ \bibinfo {pages} {24}
  (\bibinfo {year} {2009})},\ \Eprint {http://arxiv.org/abs/0903.2275}
  {arXiv:0903.2275 [astro-ph.HE]} \BibitemShut {NoStop}%
\bibitem [{\citenamefont {{Rieger}}\ and\ \citenamefont
  {{Duffy}}(2006)}]{rieger06}%
  \BibitemOpen
  \bibfield  {author} {\bibinfo {author} {\bibfnamefont {F.~M.}\ \bibnamefont
  {{Rieger}}}\ and\ \bibinfo {author} {\bibfnamefont {P.}~\bibnamefont
  {{Duffy}}},\ }\href {\doibase 10.1086/508056} {\bibfield  {journal} {\bibinfo
   {journal} {\apj}\ }\textbf {\bibinfo {volume} {652}},\ \bibinfo {pages}
  {1044} (\bibinfo {year} {2006})},\ \Eprint
  {http://arxiv.org/abs/arXiv:astro-ph/0610187} {arXiv:astro-ph/0610187}
  \BibitemShut {NoStop}%
\end{thebibliography}

%

\end{document}